\documentclass[usegraphicx,usenatbib]{mn2e}
\usepackage{amssymb}
\usepackage{graphicx}
\usepackage{longtable}

\def\Teff{$T_{\rm eff}$}
\def\logg{$\log\,g$}

\def\Vt{V${\rm t}$}

\newcommand {\apgt} {\ {\raise-.5ex\hbox{$\buildrel>\over\sim$}}\ }
\newcommand {\aplt} {\ {\raise-.5ex\hbox{$\buildrel<\over\sim$}}\ }

\title[Mn abundances in the disc stars ]
{Mn abundances in the stars of the Galactic disc with metallicities 
--1.0 $<$ [Fe/H] $<$ 0.3
\thanks{Based on observations collected at OHP observatory, France}
\thanks{Table 1  are only available in electronic form}
}
\author[T.~Mishenina  et al.]
{T.~Mishenina$^{1}$,
 T.~Gorbaneva$^{1}$,
M.~Pignatari$^{2,3,4}$,
F.-K.~Thielemann$^{3}$,\and
S.A.~Korotin$^{1}$ \\
$^{1}$Astronomical Observatory, Odessa National University,         and \\
    Isaac Newton Institute of Chile, Odessa branch,
       Shevchenko Park, 65014, Odessa, Ukraine\\
$^{2}$ Konkoly Observatory, Research Centre for Astronomy and Earth Sciences, 
Hungarian Academy of Sciences, \\
       Konkoly Thege Miklos ut 15-17, H-1121 Budapest, Hungary\\
$^{3}$ Department of Physics, University of Basel, Klingelbergstrabe 82,
        4056 Basel, Switzerland\\
$^{4}$ The NuGrid collaboration, http://www.nugridstars.org\\ }

\begin{document}

\date{Accepted 2015 xxx. Received 2015 xxx; in original form 2015 xxx}
\pagerange{\pageref{firstpage}--\pageref{lastpage}}
\pubyear{2015}

\maketitle

\label{firstpage}

\begin{abstract}
In this work we present and discuss the observations of the Mn abundances for 
247 FGK dwarfs, located in the Galactic disc with metallicity 
-$1<$[Fe/H]$<+0.3$. The observed stars belong to the substructures of the 
Galaxy thick and thin discs, and to the Hercules stream.
The observations were conducted using the 1.93 m telescope at Observatoire de 
Haute-Provence (OHP, France) equipped with the echelle type spectrographs 
ELODIE and SOPHIE. The abundances were derived under the LTE approximation, 
with an average error for the [Mn/Fe] ratio of 0.10 dex.
For most of the stars in the sample Mn abundances are not available in the 
literature. We obtain an evolution of [Mn/Fe] ratio with the metallicity [Fe/H] 
consistent with previous data compilations. In particular, within the 
metallicity range covered by our stellar sample the [Mn/Fe] ratio is increasing 
with the increase of metallicity. This due to the contribution to the Galactic
chemical evolution of Mn and Fe from thermonuclear supernovae.
We confirm the baseline scenario where most of the Mn in the Galactic disc and 
in the Sun is made by thermonuclear supernovae. In particular, the effective 
contribution from core-collapse supernovae to the Mn in the Solar system is 
about 10-20\%. However, present uncertainties affecting the production of Mn 
and Fe in thermonuclear supernovae are limiting the constraining power of the 
observed [Mn/Fe] trend in the Galactic discs on, e.g., the frequency of 
different thermonuclear supernovae populations. The different production of 
these two elements in different types of thermonuclear supernovae needs to be 
disentangled by the dependence of their relative production on the metallicity 
of the supernova progenitor.
\end{abstract}


\section{Introduction}
Manganese (Mn, Z=25) is a monoisotopic element member of the iron group. In 
stellar spectra several Mn absorption lines are known. Since early studies of 
stellar chemical composition, it was observed that in metal-poor stars Mn has a 
different behaviour with respect to Fe compared to other iron-peak elements. 
\citep[e.g.,][]{wallerstein:62}.
The chemical evolution of [Mn/Fe] is also different compared to 
$\alpha$--elements (O, Mg, Si, S, Ca and Ti), which abundances increase with 
the metallicity decreasing \citep[e.g.,][]{gratton:89}. 

Today a large number of Mn spectroscopic observations are available for stars 
with different age and metallicity and from different galactic hosts, e.g., 
from our Galaxy, from Globular Clusters (GCs) including Omega-Cen, and from 
Dwarf Spheroidal Galaxies \citep[e.g.,][]{prochaska:00, mcwilliam:03, 
alves-brito:07,sobeck:06,cunha:10,pancino:11}.
In particular, the Mn abundance survey in globular clusters and field stars 
within the metallicity range $-2.7$ $<$ [Fe/H] $<$ $-0.7$ by \cite{sobeck:06} 
found consistent average [Mn/Fe] ratios ([Mn/Fe]=$-0.36$ and $-0.37$, 
respectively) for those two populations.
This is consistent with the fact that for metallicities [Fe/H]~$\lesssim-1$ 
core-collapse supernovae (CCSNe) are the only astrophysical producers of Mn in 
the Galaxy \citep[e.g.,][]{thielemann:96,woosley:02,nomoto:13}. The same is 
true for GCs, where the mass of the cluster is not large enough to keep the 
supernovae ejecta, and all Mn and Fe observed is due to pollution from massive 
stars before the GC formed. This is the main reason for the good agreement for 
the average [Mn/Fe] observed in unevolved halo stars and in GCs 
\citep[][]{sobeck:06}.
On the other hand, the star-to-star scatter reported by \cite{sobeck:06} is 
$-0.6 \lesssim$~[Mn/Fe]~$\lesssim$~0, that is larger than the reported 
observational errors. The same conclusion may be derived by the observation in 
halo stars ($-1.1 \lesssim$~[Mn/Fe\rbrack~$\lesssim 0.5$) and from Omega Cen 
\citep[$-0.8 <$~[Mn/Fe\rbrack~$< -0.2$,][]{cunha:10, pancino:11}.
Such a large spread is difficult to reconcile with baseline one-dimensional 
CCSN models \citep[e.g.,][]{woosley:95,limongi:00}, where for an amount of 
$^{56}$Ni ejected in the order of 0.1 M$_{\odot}$ the stellar yields show an 
[Mn/Fe] ratio $\gtrsim -0.5$. 
Therefore, the origin of the [Mn/Fe] in the range 
$-1.0 \lesssim$~[Mn/Fe]~$\lesssim -0.5$ is not clearly understood \citep[][]{andrievsky:07},
although an increase of Ye in the ejecta  \citep[see Fig.5 in][]{thielemann:96}, as expected from  
neutrino interactions  \citep[see][]{frohlich:06} can explain smaller values. 
An alternative scenario, in order to reproduce the low [Mn/Fe] 
ratio is the contribution from Hypernovae, ejecting large quantities of Fe 
compared to Mn \citep[e.g.,][]{umeda:05,nomoto:06,nomoto:13}.

The observation of Mn abundances may be affected by deviations from the LTE. 
Bergemann \& Gehren (2008) have found that [Mn/Fe] ratio measured in LTE 
approximation might underestimate the real Mn abundances up to 0.5--0.6 dex 
for the metal-poor stars, and $\sim$ 0.1 dex for stars of solar-like 
metallicities .

For metallicities higher than [Fe/H]$\sim-1$ typical of the Galactic discs, 
the observation of the [Mn/Fe] ratio is still controversial. \cite{nissen:00} and \cite{reddy:06} 
reported similar Mn abundance trends with [Fe/H] for thick 
and thin disc stars. On the other hand, \cite{feltzing:07} and \cite{battistini:15} found a different behaviour in the two stellar populations. 
A similar result is obtained by \cite{hawkins:15} for giant stars. 

The nucleosynthesis of Mn becomes more complex for metallicities typical of the 
Galactic disc. Thermonuclear supernovae 
\citep[SNIa, e.g.,][for a recent review]{hillebrandt:13} start to contribute to 
the production of Mn and Fe for [Fe/H]$\gtrsim-1$ \citep[][]{matteucci:86}, 
leading to the observed increasing trend of [Mn/Fe] up to the present solar 
values. In particular, most of the Fe and Mn observed today in the Solar system 
are made by SNe Ia \citep[e.g.,][]{timmes:95,cescutti:08,kobayashi:11}. The 
production of Fe in SNe Ia as $^{56}$Ni is not completely independent from the 
initial metallicity of the star, and its production tends to decrease with the 
increasing metallicity \citep[][]{timmes:03,travaglio:05,bravo:10}. On the 
other hand, Mn production increases with the initial metallicity of the SNIa 
progenitor \citep[e.g.,][]{iwamoto:99, travaglio:05}, which is consistent with the 
increasing [Mn/Fe] with [Fe/H]. The metallicity dependence of Mn production in 
SNe Ia was also inferred by Galactic chemical evolution (GCE) simulations 
\citep[][]{cescutti:08}.

One of the main uncertainties affecting the evolution of Mn abundance in the 
Galactic disc is our present understanding of SNIa nucleosynthesis. The two 
historical scenarios proposed for SNe Ia are the single-degenerate scenario 
(SDS), where an accreting white dwarf (WD) is reaching the Chandrasekhar mass 
by accreting material from a stellar binary companion, or the double-degenerate 
scenario (DDS), where a SNIa is formed by a merger of two CO WDs. Based on 
different motivations, in the last decades one or the other scenario have been 
favoured \citep[see discussion in][]{hillebrandt:13}. Another scenario related 
to the SDS is the double-detonation SN, where the SNIa explosion is triggered 
by the He detonation initiated in the external He shell during the accretion on 
a sub-Chandrasekhar progenitor \citep[][]{pakmor:10,fink:10}.

Based on GCE calculations of the [Mn/Fe] ratio in the Galactic disc, 
\cite{matteucci:09} proposed that two different types of SNIa contributors are 
needed to reproduce the observations. More recently, based on the use of 
stellar yields from multi-dimensional hydrodynamics simulations of SNe Ia, 
\cite{seitenzahl:13} confirmed that both SNIa scenarios with a mass close to the 
Chandrasekhar limit and with sub-Chandrasekhar mass (i.e., DDS or 
double-detonation in an accreting sub-Chandrasekhar progenitor) are needed to 
fit the [Mn/Fe] evolution. In particular, the yields of the first ones carry a 
much larger Mn/Fe ratio than the second ones.

In this work we provide the measurement for Mn abundances for 247 F-G-K-type 
dwarf stars located in the thin and thick disc populations, and for the 
Hercules Stream, covering the metallicity range $-1.0 <$ [Fe/H] $<$ 0.3.
The main purpose of this survey is to provide new constraints helping to 
definitely establish the origin of the Mn production in the Galactic disc.
The paper is organized as follow. 
The observations and selection of stars, and definition of the main stellar 
parameters are described in \S \ref{sec: stellar param}. The abundance 
determinations for Mn and the error analysis are presented in 
\S \ref{sec: mn_abundance}. In \S \ref{sec: results_and_comparison} the results 
are compared with other measurements available in the literature. The 
implications of the results and the nucleosynthesis of Mn in stars is reported 
in \S \ref{sec: mn_nucleosynthesis}. Conclusions are drawn in 
\S \ref{sec: conclusions}.

\section{Observations, selection and parameters of the disc stars}
\label{sec: stellar param}

The investigated stellar spectra were obtained using the 1.93 m telescope at 
Observatoire de Haute-Provence (OHP, France) equipped with  echelle type 
spectrographs, namely SOPHIE, resolving power R = 75 000 
\citep{perruchot:08} and ELODIE, R = 42 000 \citep{baranne:96}, in the 
wavelength range  $\lambda$~4400--6800\,\AA\, and signal-to-noise ratio of 
about 100-300. 

\noindent
To distinguish between stars in the thin disc, thick disc and the Hercules 
Stream, we considered the probability that a star belongs to either one of them 
accounting for its spatial velocity, kinematic disc parameters, and the 
quantity and percentage of every disc stars in our sample. The stars that 
belong to the Galactic sub-structures were selected using the technique 
described in \cite{mishenina:04}. The primary processing of spectra was carried 
out immediately during observations \citep{katz:98}. Further spectra 
processing such as the continuum placement, line depth and equivalent width (EW) 
measurements, etc., was conducted using the DECH20 software package by 
Galazutdinov \citep{galazut:92}. The atmospheric parameters of target stars 
has been determined earlier. The methods applied are described in detail in 
\cite{mishenina:01}, \cite{mishenina:04} and \cite{mishenina:08}.

The effective temperatures \Teff\ are defined by calibration of line-depth 
ratios (R1/R2) for spectral line pairs that markedly differ in their low-level 
excitation potential \cite{kovtyukh:03}. A large number of calibrations 
(80--103) permitted to reduce the influence of errors in the line-depth and 
atmospheric parameter measurements on the resulting temperature estimates. The 
intrinsic accuracy of the method applied for dwarfs is 5--45 K. For the stars 
with [Fe/H] $<-0.5$ the \Teff\ were estimated by adjustment of far-wings of 
the $\rm H_{\alpha}$ line \citep[][]{mishenina:01}. 

The surface gravities \logg\ values  are computed by the ionization balance of the 
neutral and ionized iron. This method implies that similar abundances are 
obtained from the neutral iron FeI and ionized iron FeII lines. Its accuracy is 
affected by a number of factors, such as uncertainty of oscillator strengths of 
the log gf lines and thermal structure of atmospheric models, and NLTE effects.

However, the determination of the surface gravity by the parallax method is 
also affected by uncertainties. In order to determine the stellar mass by using 
theoretical evolution tracks, it is necessary to measure in advance the 
metallicity and $\alpha$--element enrichment, leading to an uncertainty in the 
mass determination of $\sim$ 0.2M$_{\odot}$, which corresponds to an error for 
the surface gravity of $\sim$ 0.1 dex. Therefore, NLTE effects and atmospheric 
model uncertainties still affect the analysis. As reported by 
\cite{allendeprieto:99}, the astrometric and spectroscopic methods provide 
consistent results in the metallicity range $-1 <$ [Fe/H] $<$ 0.3. Having 
compared the resulting \logg\ obtained by us using the ionization balance 
method with the \logg\ value computed by using the parallax in 
\cite{allendeprieto:99} for 39 stars in common, leads to differences 
 not exceeding 0.1 dex on average \cite{mishenina:04}. 

The microturbulence velocity \Vt\ is derived considering that the iron abundance 
obtained from the given FeI line is not correlated with the equivalent width EW 
of that line. The adopted value of metallicity [Fe/H] is the iron abundance 
derived from the FeI lines. The determination errors are: for the effective 
temperatures $\delta$\Teff $= \pm 100$K, the surface gravities 
$\delta$\logg $= \pm0.2$dex, the microturbulence velocity 
$\delta$\Vt $= \pm 0.2$km/sec. The obtained parameter values and their 
comparison with the results of other authors are reported in 
\cite{mishenina:04} and \cite{mishenina:13}.

\section{The Mn abundance}
\label{sec: mn_abundance}

\begin{figure}
\begin{tabular}{c}
\includegraphics[width=8cm]{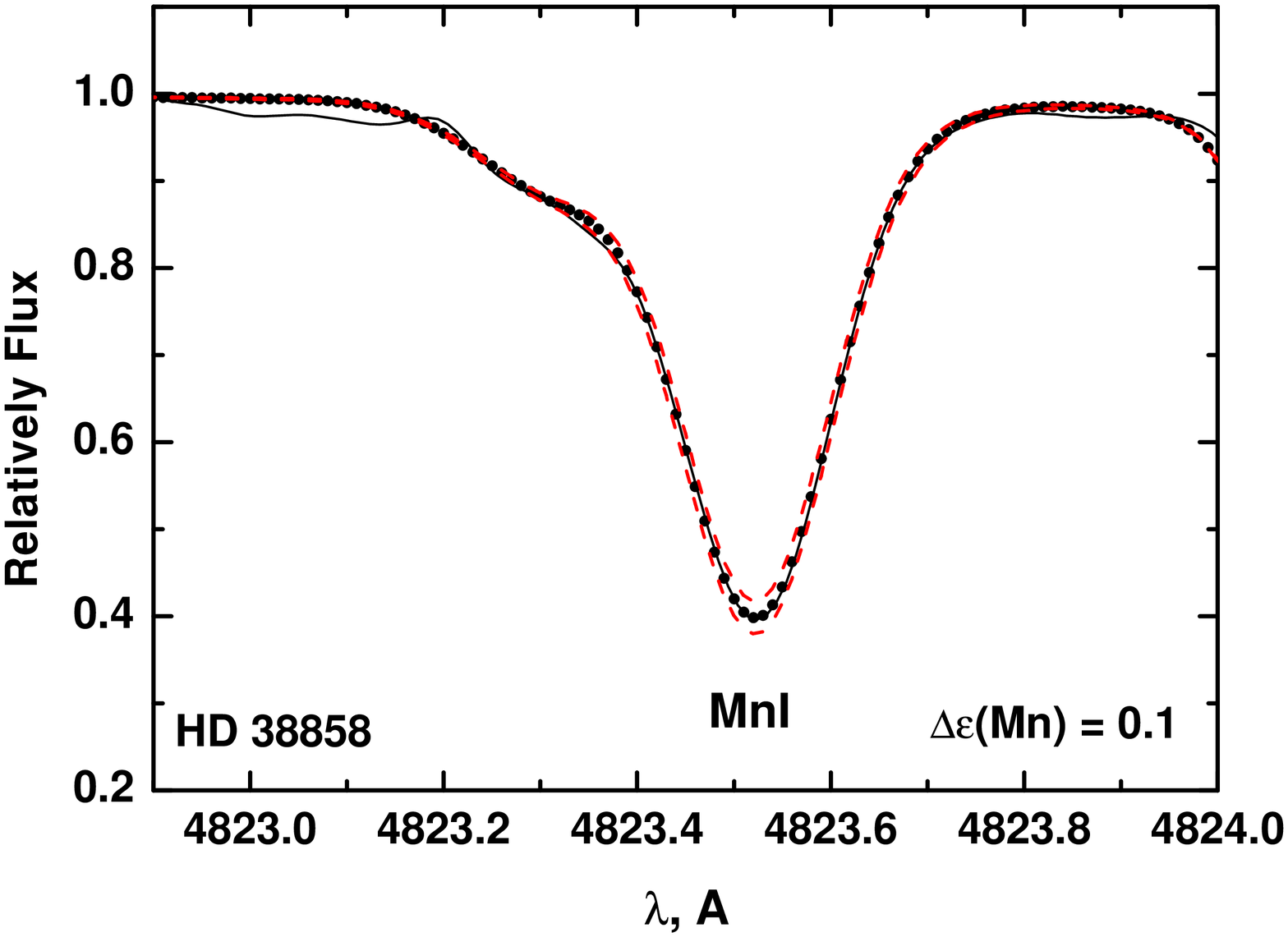}\\
\includegraphics[width=8cm]{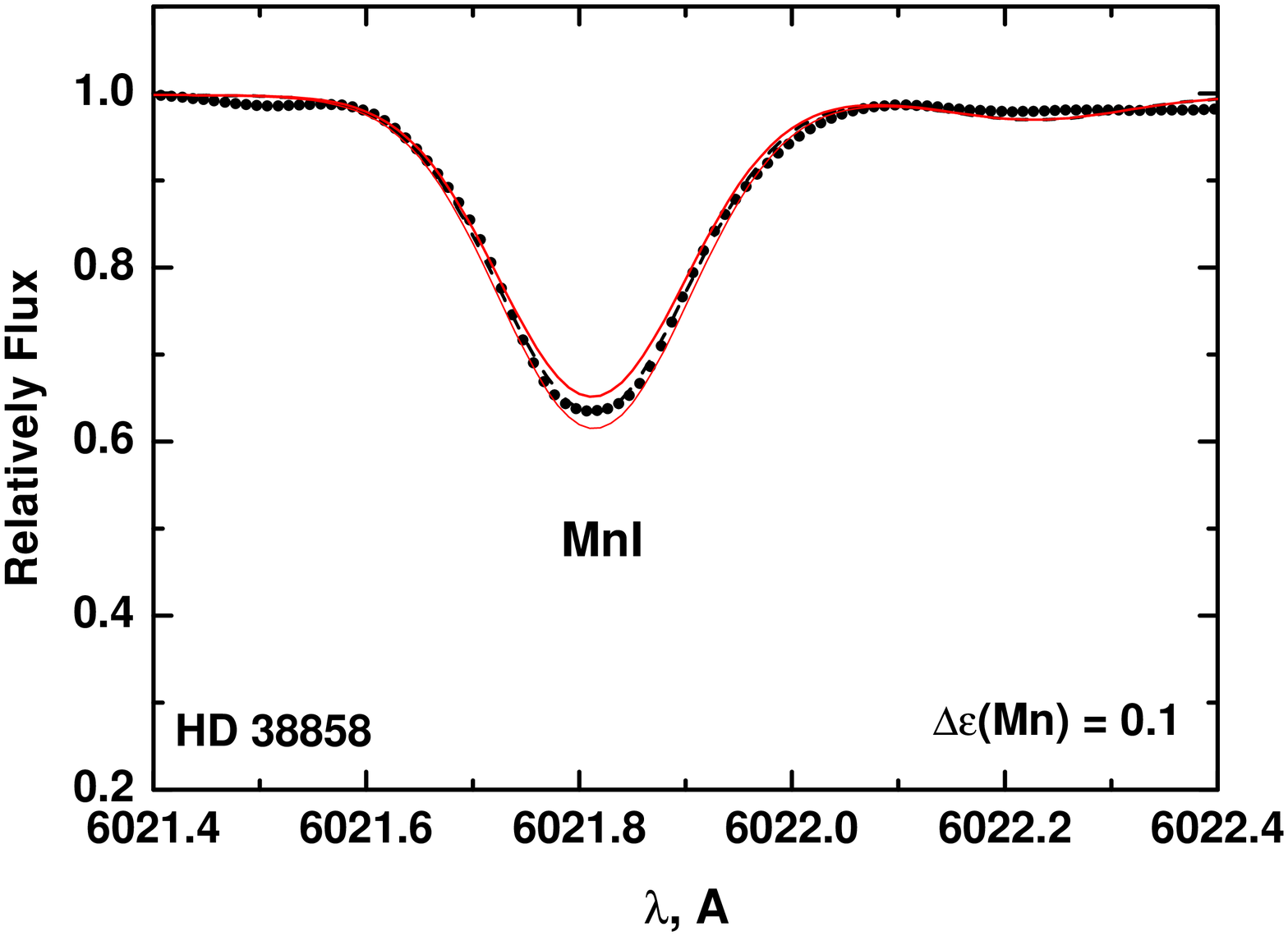}\\
\end{tabular}
\caption{Example of the fitting
of the observed spectrum (black dots with solid line) by the synthetic spectrum
(red dotted line) in the area of the Mn lines.}
\label{prof}
\end{figure}

The Mn abundances were derived by computing the synthetic spectrum in 
the region of the Mn lines by the newly updated STARSP LTE software 
\citep[][]{tsymbal:96} using the Kurucz models \citep[][]{castelli:04}. The MnI 
lines undergo hyperfine-structure (HFS) splitting, due to the interaction of 
the magnetic moment of the nucleus with the magnetic field of the electrons. 
The list of lines and HFS data were taken from \cite{prochaska:00}. The van der 
Waals damping constant C6 was taken from \cite{bergemann:08}. Atomic data for 
other lines required to compute the synthetic spectrum in the region of the MnI 
lines were taken from the Vienna Atomic Line Database VALD \citep[][]{kupka:99}.
The solar MnI abundances are derived for each line by the solar spectra 
reflected by the moon and asteroids. Two echelle type spectrographs were adopted.
In order to select reliable fitting lines, we started with derivation of the Mn 
abundances from 16 MnI lines in the atmospheres of the Sun and some other stars,
using spectra obtained with both spectrographs. A total of 6 solar spectra were 
used, including three spectra obtained with the echelle type spectrograph ELODIE
and three spectra received by the echelle type spectrograph SOPHIE. Five lines 
were obtained: 4783, 4823, 5432, 6013 and 6021~\AA. The equivalent widths EWs 
averaged by spectra for each spectrograph differ by 1--2\% for strong lines and 
2--3\% for the 5432~\AA\ line.  The mean difference of EWs of Mn lines in spectra
obtained with two spectrographs is 
$<$(EWs(Mn)$_{SOPHIE}$ - EWs(Mn)$_{ELODIE}$)$>$ = --1.8 $\pm$ 0.7~m\AA.~ 
Further, we also compared of EWs of Fe I  lines  measured in solar spectra obtained with two spectrographs.
 The mean difference in this case is 
$<$(EWs(Fe)$_{SOPHIE}$ - EWs(Fe)$_{ELODIE}$)$>$ = --1.2 $\pm$ 2.3~m\AA.~      
Hence, the MnI abundance derived from selected lines in the solar spectra obtained 
with both spectrographs are consistent. Lines 4783 and 4823~\AA\ are rather 
strong, so they were not used in the analysis of cooler stars (\Teff\ $<$ 
5200 K) and those more enriched in iron ([Fe/H] $>$ 0). For each line we 
derived the Mn abundances in the solar and stellar spectra, by fitting locally 
the observed spectra with the synthetic model. For a given line we compare the 
abundance value to the solar abundance. We adopted this differential approach 
to eliminate the impact of potential errors in the oscillator strengths led to the following 
adopted atomic parameters. Finally, we derived the stellar Mn abundance by 
averaging the values obtained for each line (Table \ref{abun}). The mean solar 
abundance computed for the lines with log gf from VALD data base 
\citep{kupka:99} and the solar model from \cite{castelli:04} and adopted values 
for each line in this work  are 5.3, 5.25, 5.27, 5.24, and 5.24 for the lines 
$\lambda$ 4783, 4823, 5432, 6013, 6021~\AA, respectively. In. Fig. \ref{prof} 
we show few examples of profile fitting for Mn lines.

\begin{table*}
\caption{Atmospheric parameters and Mn abundance  log A(Mn) each used line 
($\lambda$, A) and [Mn/Fe] ratio for our target stars}
\label{abun}
\begin{tabular}{lccrccccccr}
\hline
\hline
Star       & \Teff,K & \logg & [Fe/H] &  \Vt   &4783~\AA&4823~\AA &5432~\AA& 6013~\AA& 6021~\AA& [Mn/Fe] \\
\hline                                                                                              
  Sun      &        &        &        &        &   5.30 &   5.25  &  5.27 &    5.24 &   5.24&        \\
Thick disc &        &        &        &        &        &         &       &         &       &        \\
HD245      &   5400 &   3.4  &  -0.84 &   0.7  &   4.32 &   4.35  &  4.10 &    4.18 &   4.18&  -0.16 \\
HD3765     &   5079 &   4.3  &   0.01 &   1.1  &        &         &  5.30 &    5.30 &   5.30&   0.04 \\
HD6582     &   5240 &   4.3  &  -0.94 &   0.7  &   4.20 &   4.20  &  4.18 &    4.15 &   4.15&  -0.14 \\
HD13783    &   5350 &   4.1  &  -0.75 &   1.1  &   4.35 &   4.40  &  4.40 &         &   4.25&  -0.17 \\
HD18757    &   5741 &   4.3  &  -0.25 &   1.0  &   4.95 &   4.95  &  4.95 &    4.85 &   4.85&  -0.10 \\
...        &   ...  &  ...   &  ...   &   ...  &   ...  &   ...   &  ...  &   ...   &   ... &   ...  \\
...        &   ...  &  ...   &  ...   &   ...  &   ...  &   ...   &  ...  &   ...   &   ... &   ...  \\
...        &   ...  &  ...   &  ...   &   ...  &   ...  &   ...   &  ...  &   ...   &   ... &   ...  \\
\hline                                                                                               
\end{tabular}                                                                                        
\end{table*}

\noindent
For a sample of stars also analysed by other studies, we compare the inferred 
atmospheric parameters in Table \ref{ncap}. Impact of these variations are 
within the observational errors for Mn abundance. For a detailed comparison we show in 
Table \ref{compar}  the stellar data for each star together with other 
works. \cite{takeda:07} and \cite{nissen:11} show the largest variations 
compared to our \Teff, but they also report larger errors 
($\sigma$ $\gtrsim$ 100 K). The biggest departure is for HD 4307 by 
\cite{takeda:07}, with a difference for \Teff~ larger than 200 K. 
In the Table \ref{comparBB} we presented the comparison of the data of \cite{battistini:15} and our determinations of parameters and Mn abundance for common stars. 
A good agreement is obtained between these two data sets.

\begin{table*}
\begin{center}
\caption[]{Comparison of our parameters and Mn abundance determinations with 
the results of other authors for the $n$ stars shared with our stellar sample. 
}
\label{ncap}
\begin{tabular}{cccccc}
\hline
 & $\Delta$(\Teff) & $\Delta$(\logg) & $\Delta$([Fe/H]) & $\Delta$([Mn/Fe]) & n \\
\hline
Feltzing et al.& 24 & -0.03 & -0.01 & 0.02 & 10 \\
 2007 &$\pm$76 & $\pm$0.13 & $\pm$0.08 & $\pm$0.06 &  \\
Reddy et al. & 105 & -0.18 & 0.01 & 0.14 & 9  \\
2006 &$\pm$100 & $\pm$0.21 & $\pm$0.11 & $\pm$0.06 &  \\
Takeda & -14 & -0.06 & -0.04 & 0.02 & 31  \\
et al. 2007  &$\pm$119& $\pm$0.21 & $\pm$0.10 & $\pm$0.08 &  \\
Nissen et al.& 7 & -0.03 & -0.05 & 0.02 & 4 \\
2011  &$\pm$143 & $\pm$0.20 & $\pm$0.10 &  $\pm$0.02 &   \\
Adibekyan & 28 & -0.07 & 0.01 & 0.01 & 9  \\
et al. 2014 &$\pm$57& $\pm$0.14 & $\pm$0.04 & $\pm$0.05 &  \\
Battistini \& Bensby& -4 & -0.10 & -0.03 & 0.02 & 22 \\
 2015 &$\pm$106 & $\pm$0.15 & $\pm$0.06 & $\pm$0.06 &  \\
\hline
\end{tabular}
\end{center}
\end{table*}

\begin{table*}
\caption{Comparison of atmospheric parameters and Mn abundance for common stars. }
\label{compar}
\begin{tabular}{llccrr}
\hline
HD    &	  sourses                &      \Teff\	  &    \logg\	&[Fe/H]&[Mn/Fe]   \\
\hline
4307   &  Adibekyan et al. 2012	 &      5812	  &      4.10	&--0.23&  --0.07    \\
      &  Takeda  2007            &      5648      &	 3.75	&--0.29&  --0.09    \\
      &      our                 &      5889	  &      4.00	&--0.18&  --0.06     \\
6582  &	   Reddy et al.  2006 	 &      5300	  &      4.67	&--0.86&  --0.23     \\
      &      Takeda  2007        &      5330	  &      4.54	&--0.81&  --0.12    \\
      &           our            &      5240	  &      4.30	&--0.94&  --0.14     \\
22879 &	Adibekyan et al. 2012	 &      5884	  &      4.52	&--0.82&  --0.30    \\
      &	Feltzing et al. 2007     &  	5920	  &      4.33	&--0.84&  --0.18     \\
      &	Nissen\&Shuster 2011     &	5759	  &      4.25	&--0.85&  --0.27     \\
      &	Reddy et al.  2006       & 	5722	  &      4.50	&--0.87&  --0.39     \\
      &          our             &      5972	  &      4.50	&--0.77&  --0.22     \\
76932 &	Feltzing et al. 2007     &      5875	  &      4.10	&--0.91&  --0.23     \\
      &	Nissen\&Shuster 2011     &	5877	  &      4.13	&--0.87&  --0.25     \\
      &	Reddy et al.  2006       & 	5783	  &      4.09	&--0.86&  --0.35     \\
      &            our           &      5840	  &      4.00	&--0.95&  --0.25     \\
106516&	Nissen\&Shuster 2011     &	6196	  &      4.42	&--0.68&  --0.23     \\
      &	Reddy et al.  2006       &  	6069	  &      4.44	&--0.74&  --0.35     \\
      &             our          &      6165	  &      4.40	&--0.72&  --0.23     \\
125184&	Adibekyan et al. 2012    &      5680	  &      4.10	&  0.27&    0.06     \\
      &	Takeda  2007             & 	5629	  &      4.02	&  0.25&    0.24     \\
      &            our           &      5695	  &      4.30	&  0.31&    0.02      \\
157214&	Reddy et al.  2006       &  	5605	  &      4.49	&--0.41&  --0.24     \\
      &	Takeda 2007              & 	5693	  &      4.21	&--0.37&  --0.15    \\
      &             our          &      5820	  &      4.50	&--0.29&  --0.11     \\
159482&	Nissen\&Shuster 2011     &	5737	  &      4.31	&--0.73&  --0.23     \\
      &	Reddy et al.  2006       & 	5630	  &      4.58	&--0.70&  --0.32     \\
      &        our           &  5620  &  4.10  &          --0.89&   --0.21    \\
199960&	Adibekyan et al. 2012    & 	5973	  &      4.39	&  0.28&    0.10      \\
      &	Feltzing et al. 2007     &  	5924	  &      4.26	&  0.28&    0.23     \\
      &	Takeda  2007             & 	5924	  &      4.26	&  0.28&  --0.04    \\
      &            our           &      5878	  &      4.20	&  0.23&    0.02     \\
217014&	Feltzing et al. 2007     &   	5789      &	 4.34	&  0.20&    0.02     \\
      &	Takeda  2007             & 	5779	  &      4.31	&  0.20&  --0.09    \\
      &               our        &      5763	  &      4.30	&  0.17&    0.00     \\
\hline   
\end{tabular}
\end{table*}

\begin{table*}
\caption{Comparison of atmospheric parameters and Mn abundance for common stars with Battistini \& Bensby (2015) (BB 2015).  }
\label{comparBB}
\begin{tabular}{rrccrrrrrr}
\hline
HIP    & HD      &   \Teff\ &	 \logg\	& [Fe/H] &[Mn/Fe] &  \Teff\ &	 \logg\	&[Fe/H]&[Mn/Fe]      \\
       &         & BB 2015& &  &  & our &    &   &      \\
\hline
6653	&8648    &   5841&	4.3&	0.22	&0.02	&  5790&	4.2	& 0.12&	 0.02	     \\
16852	&22484	 &   6036&	4.1&	-0.03	&-0.07	&  6037&	4.1	&-0.03&	-0.04	     \\
17147	&22879	 &   5970&	4.5&	-0.81	&-0.3	&  5972&	4.5	&-0.77&	-0.22	    \\
22263	&30495	 &   5790&	4.5&	0.02	&-0.18	&  5820&	4.4	&-0.05&	-0.03	  \\
30545	&45067	 &   6042&	3.9&	-0.03	& -	&  6058&	4	&-0.02&	-0.09	    \\
38625	&64606	 &   5188&	4.4&	-0.91	&-0.08	&  5250&	4.2	&-0.91&	-0.11	      \\
38750	&64815	 &   5763&	3.9&	-0.35	&-0.19	&  5864&	4	&-0.33&	-0.14	   \\
44075	&76932	 &   5937&	4.2&	-0.9	&-	&  5840&	4	&-0.95&	-0.24	  \\
64792	&115383	 &   6185&	4.3&	0.25	&0.02	&  6012&	4.3	&0.11&	 0.00	  \\
74537	&135204	 &   5200&	4.4&	-0.19	&-	&  5413&	4	&-0.16&	-0.03	  \\
81300	&149661	 &   5216&	4.6&	-0.01	&-	&  5294&	4.5	&-0.04&	-0.03	      \\
82588	&152391	 &   5322&	4.5&	-0.08	&-	&  5495&	4.3	&-0.08&	 0.05	  \\
84905	&157089	 &   5915&	4.3&	-0.5	&-	&  5785&	4	&-0.56&	-0.19	  \\
86013	&159482	 &   5760&	4.3&	-0.81	&-0.18	&  5620&	4.1	&-0.89&	-0.21	  \\
86193	&159909	 &   5671&	4.3&	0.03	&-0.07	&  5749&	4.1	&0.06&	 0.00	      \\
88622	&165401	 &   5794&	4.5&	-0.4	&-0.16	&  5877&	4.3	&-0.36&	-0.09	  \\
93966	&178428	 &   5656&	4.2&	0.15	&-0.04	&  5695&	4.4	&0.14&	 0.00	    \\
97779	&187897	 &   5944&	4.5&	0.12	&-0.03	&  5887&	4.3	&0.08&	-0.02	    \\
98767	&190360	 &   5572&	4.5&	0.26	&-	&  5606&	4.4	&0.12&	 0.05	  \\
103682	&199960	 &   6023&	4.4&	0.33	&0.08	&  5878&	4.2	&0.23&	 0.02	  \\
104659	&201891	 &   5973&	4.3&	-1.08	&-0.26	&  5850&	4.4	&-0.96&	-0.28	  \\
113357	&217014	 &   5858&	4.4&	0.24	&0.02	&  5763&	4.3	&0.17&	0.00	  \\
\hline   
\end{tabular}
\end{table*}

\subsection{Errors in abundance determinations}

To determine the systematic errors in the Mn abundance resulting from 
uncertainties in the atmospheric parameter determinations, we derived the Mn 
abundance for several models with modified parameters 
($\delta$\Teff $= \pm100~K$, $\delta$\logg $= \pm0.2$, $\delta$\Vt $= \pm0.1$). 
The Mn abundance variations with the modified parameters and the fitting errors
for the computed and observed spectral line profiles (0.03 dex), are given in 
Table \ref{errors}.
The largest error occurs when the \Teff\ are determined inaccurately. Errors 
caused by uncertainties from other parameters are marginal. The total error
associated with the determination of the Mn abundances is 0.08 -- 0.10 dex 
(Table \ref{errors}.)

\begin{figure}
\begin{tabular}{c}
\includegraphics[width=8cm]{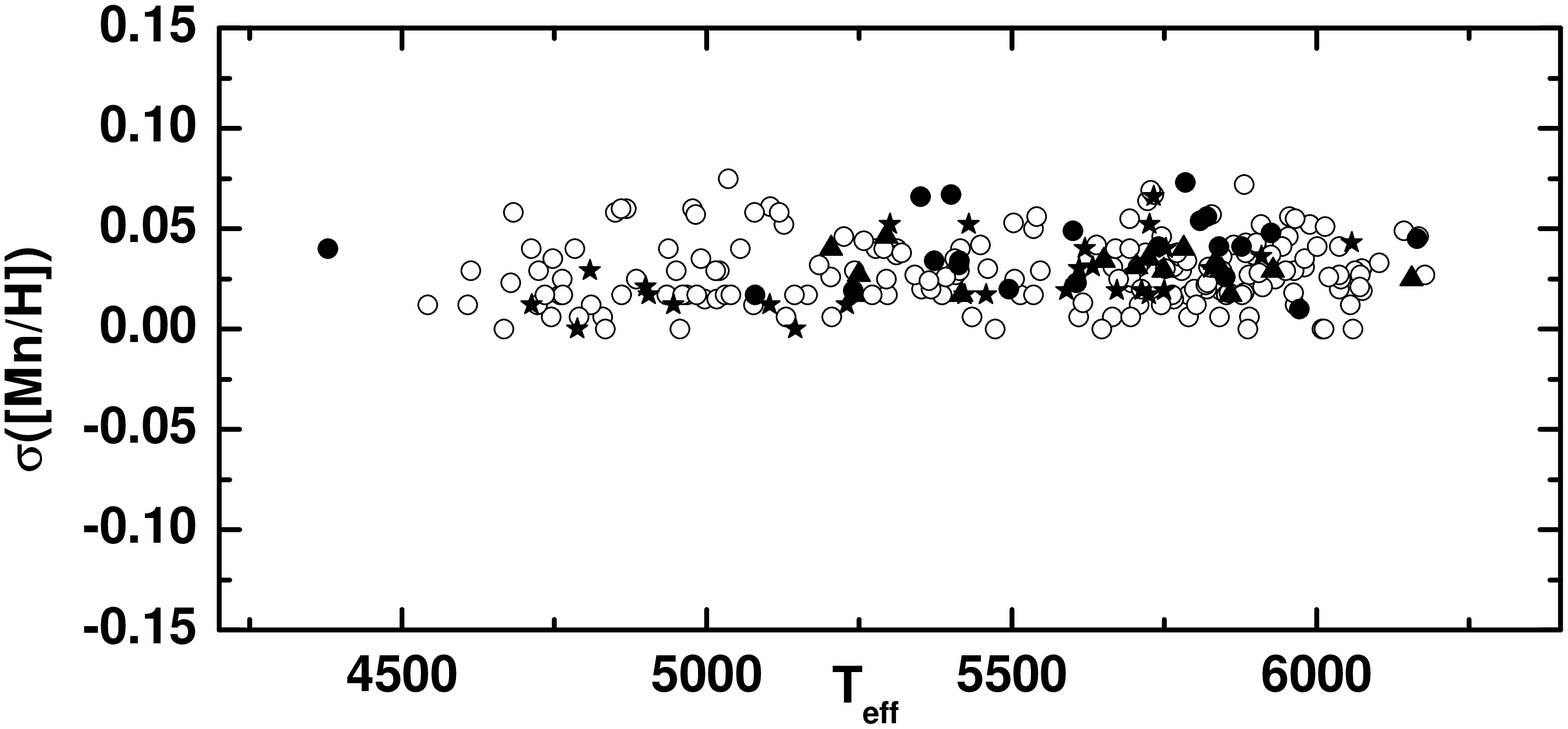}\\
\includegraphics[width=8cm]{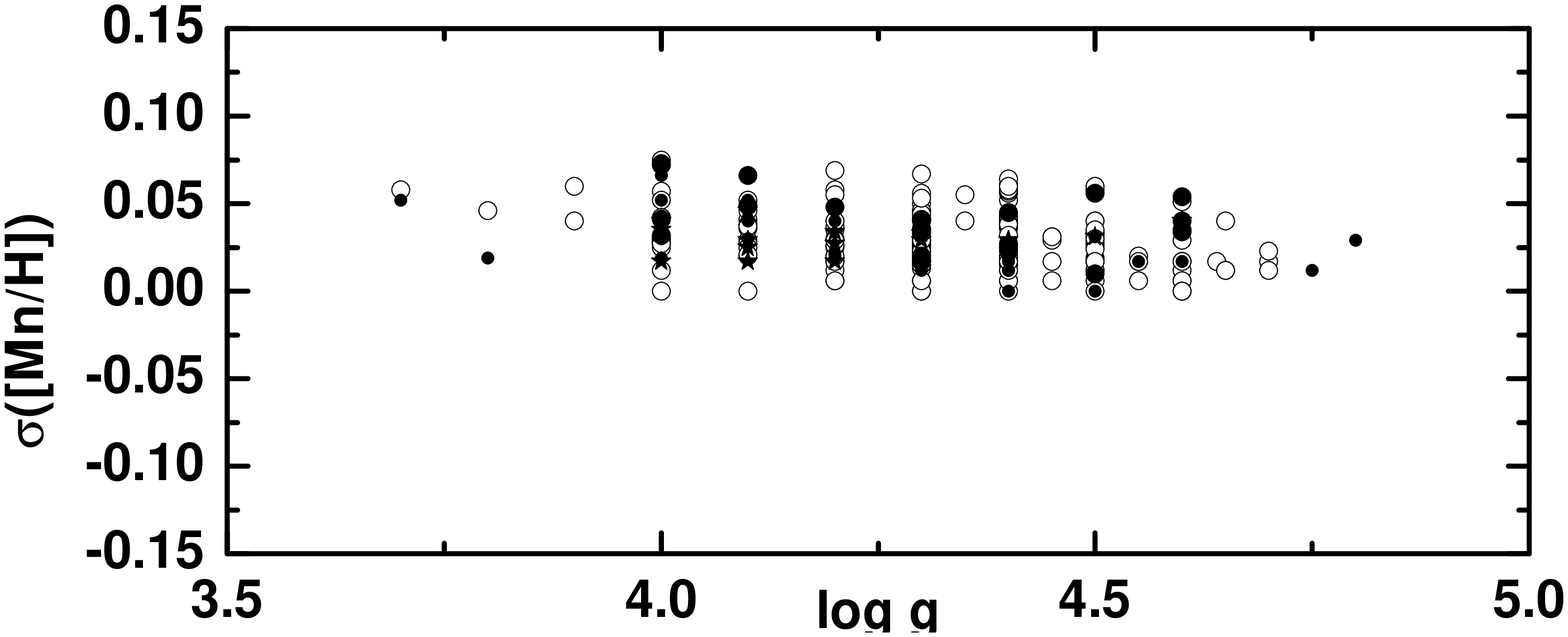}\\
\includegraphics[width=8cm]{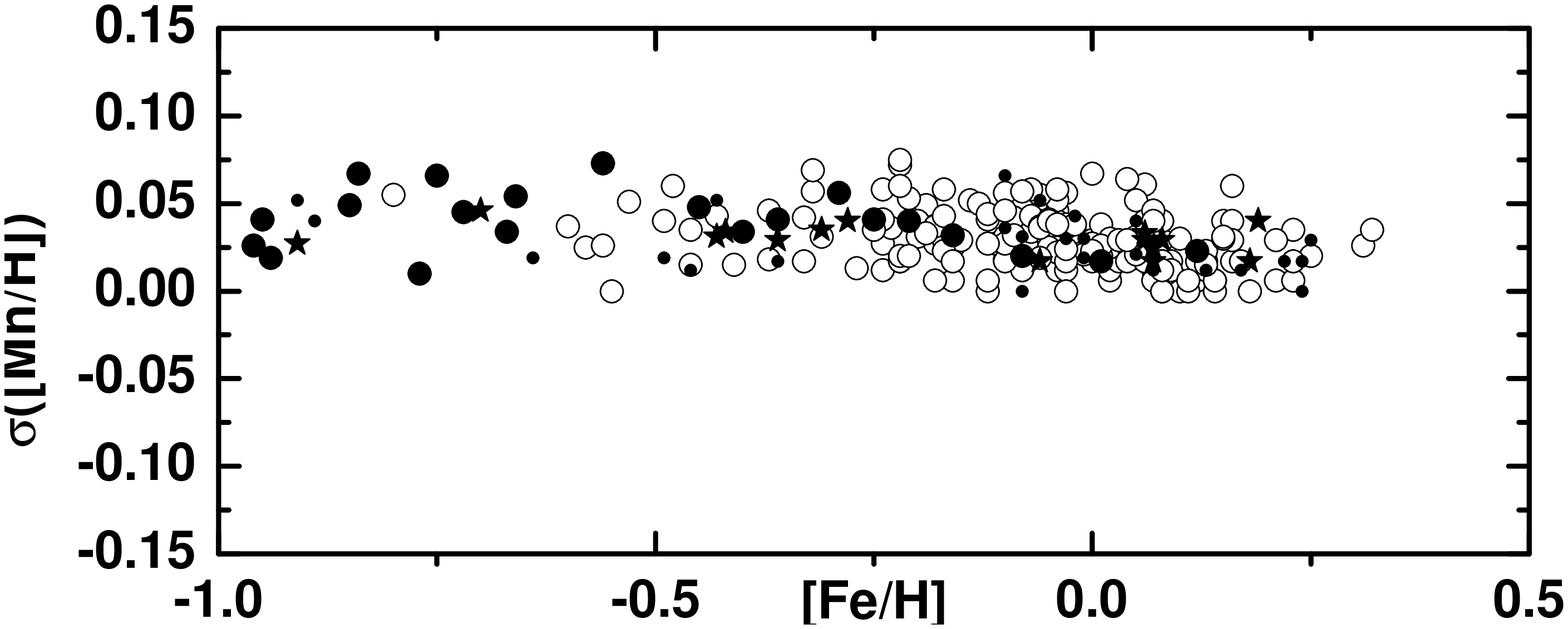}\\
\includegraphics[width=8cm]{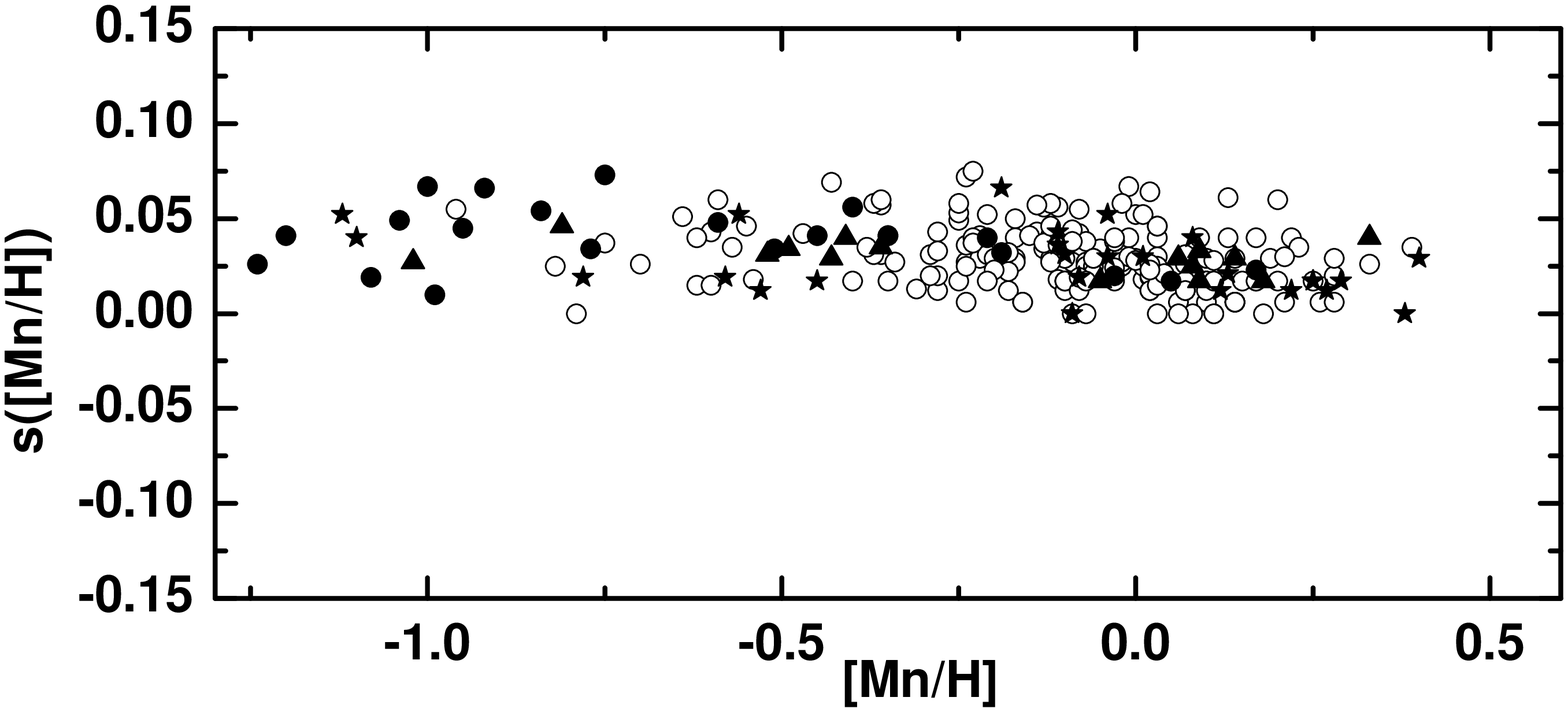}\\
\end{tabular}
\caption{Dependences of $\sigma$[Mn/H] on atmospheric parameters,  [Fe/H] and [Mn/H].}
\label{ewr7}
\end{figure}

The graph of plotted points corresponding to the Mn abundances for each line in 
spectra of all investigated stars is presented in Fig. \ref{ewr7}. In 
particular, there is no systematic difference in abundances obtained for a 
given line. The correlation between $\sigma$([Mn/H]) and [Mn/H], where 
$\sigma$ is the standard deviation, the mean value of which is 0.03, is shown 
in Fig. \ref{ewr7}. No trend is observed on the graph of $\sigma$[Mn/H] 
dependence on \Teff, \logg\ or [Fe/H] either (Fig. \ref{ewr7}).

\begin{table}
\caption{Abundance uncertainties due to atmospheric parameters.}
 \label{errors}
\begin{tabular}{lcccc}
\hline
Mn I lines  &   $\Delta$ \Teff+  & $\Delta$ \logg+ & $\Delta$ \Vt+ & tot+ \\
\hline
&HD 22879  &(5972/4.5/1.1/--0.77)& \\
\hline   
 4783	&--0.08&0.02&	0.03&         \\		     
 4823	&--0.08&0.01&	0.02&         \\		     
 5432	& -    &  - &    -  &         \\   				
 6013	&--0.06 &0.02&	0.02&		 \\    
 6021	&--0.06&0.02&	0.01&		   \\   
 average&--0.07&0.017&	0.02&		0.08  \\
\hline
  &HD 26923   & (5920/4.4/1.0/--0.03)&  \\                                                
\hline	
 4783	&--0.06&	0.03&	0.03&		\\      
 4823	&--0.07&	0.01&	0.02&	\\	      
 5432	&--0.08&	0.02&	0.03&		\\      
 6013	&--0.05&	0.02&	0.03&		  \\    
 6021	&--0.06&	0.02&	0.01&	            \\  
 average&--0.06&	0.02&	0.02&	0.08         \\
\hline   
 &HD 4635 & (5103/4.4/0.8/0.07)&  \\                                                
\hline	
 4783	&--0.10&	0.04&	0.02&		\\      
 4823	&--0.10&	0.05&	0.02&	\\	      
 5432	&--0.08&	0.03&	0.06&		\\      
 6013	&--0.06&	0.05&	0.02&		  \\    
 6021	&--0.04&	0.05&	0.04&	            \\  
 average&--0.08&	0.04&	0.03&	0.10         \\
\hline                                                                                           
\end{tabular}
\end{table}

\subsection{Analysis of the Mn spectral line parameters and evaluation of the 
effects of deviations from LTE on determination of the Mn abundance.}

\begin{figure} 
\begin{tabular}{c}
\includegraphics[width=6cm]{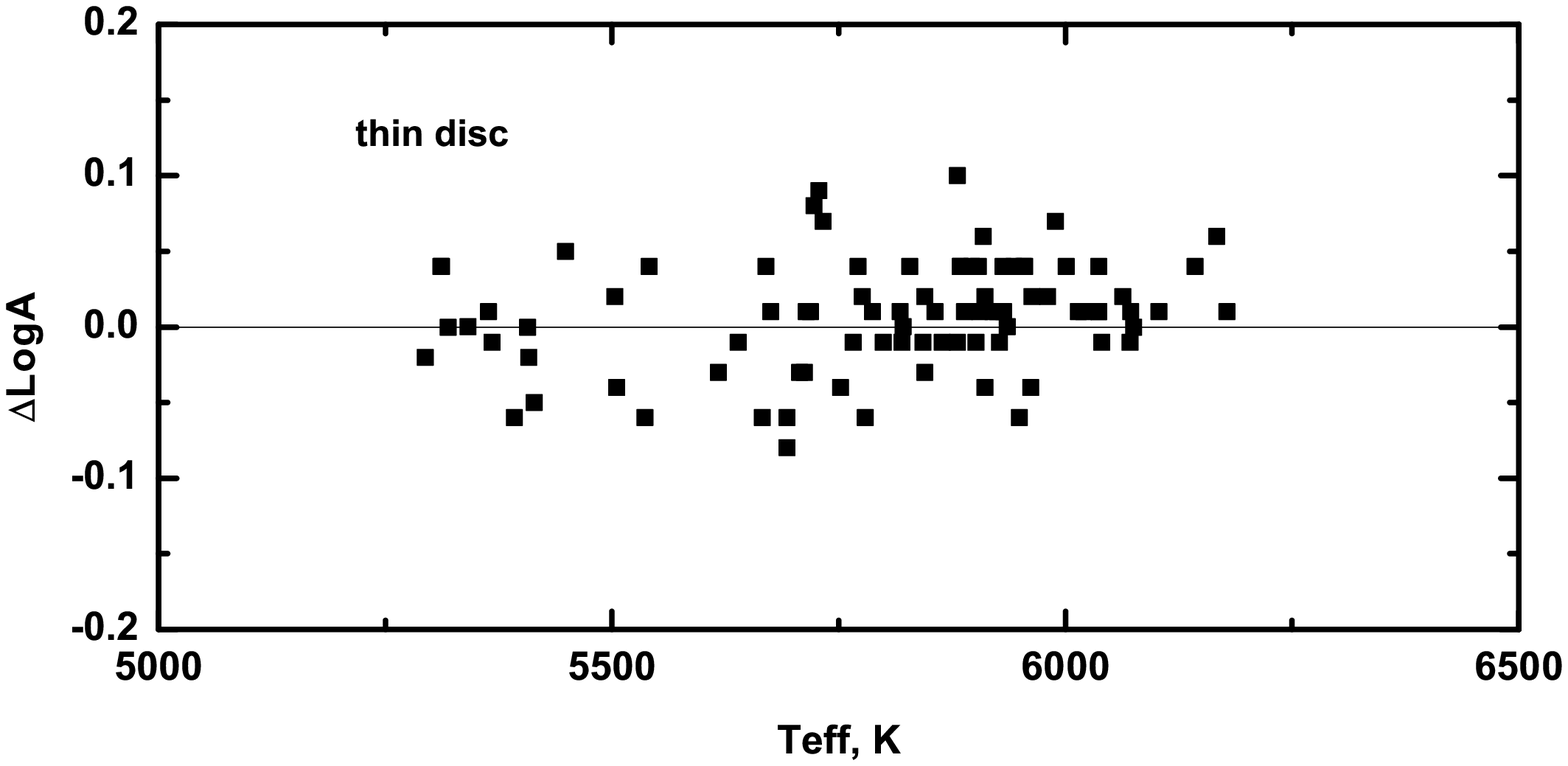}\\ \includegraphics[width=6cm]{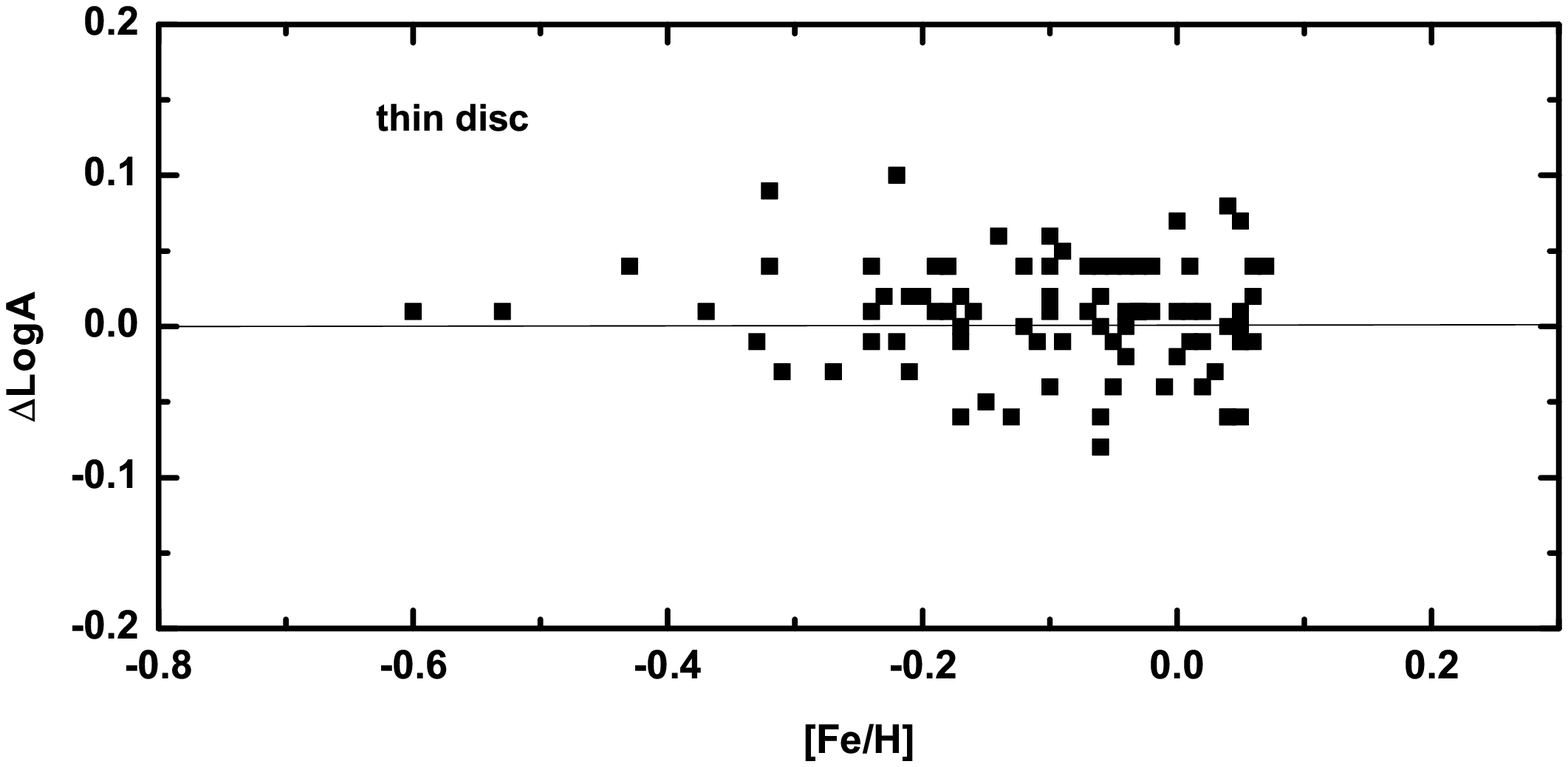}\\
\end{tabular}
\caption{
The dependence of the difference between Mn abundances determined with the Mn I lines at 4783 and 6013 AA\  for thin disc stars.
}
\label{el_fe1}
\end{figure}

\begin{figure} 
\begin{tabular}{c}
\includegraphics[width=6cm]{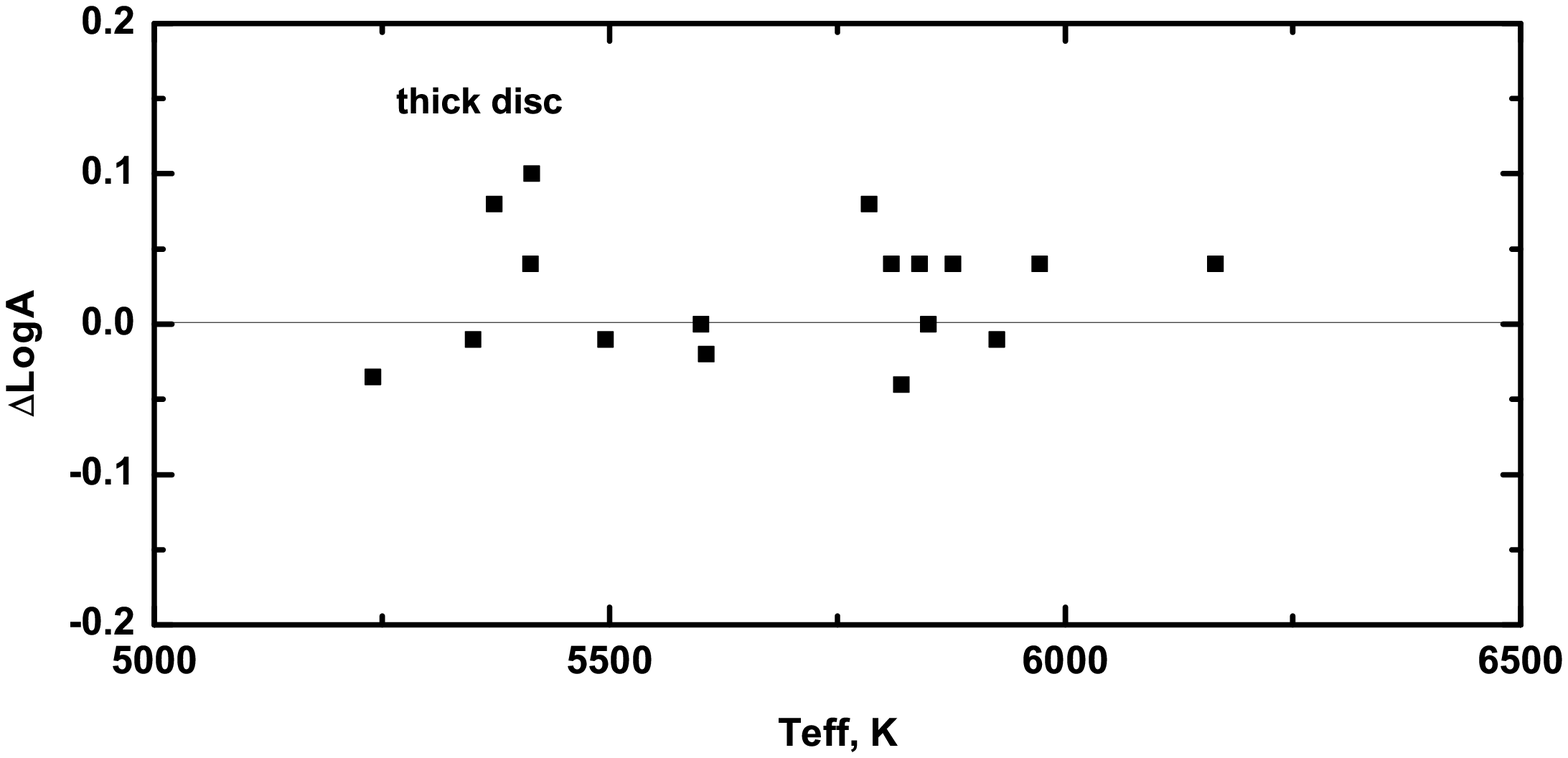}\\ \includegraphics[width=6cm]{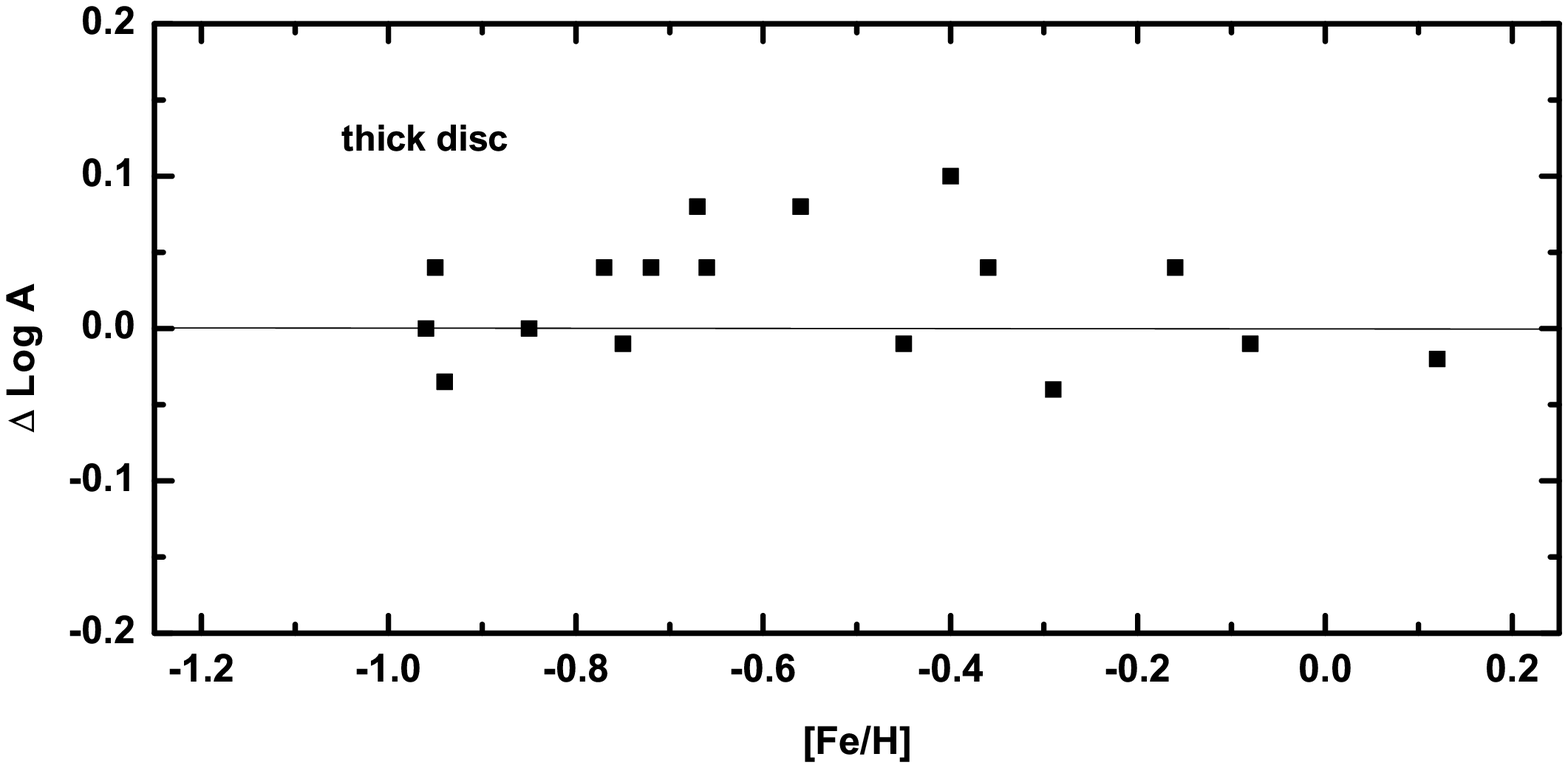}\\
\end{tabular}
\caption{
The dependence of the difference between Mn abundances determined with the Mn I lines at 4783 and 6013 AA\  for thick disc stars.
}
\label{el_fe2}
\end{figure}

In order to analyse and compare consistently the Mn abundance obtained in 
different works, it is necessary to take into account that the authors used 
different oscillator strengths log gf and different atomic data to account for 
HFS in their analysis. The values of the NLTE corrections for the Mn lines, 
including those used in our study, are investigated by \cite{bergemann:07} 
and \cite{bergemann:08}. The difference in the NLTE corrections change for 
different lines and depends on temperature and metallicity 
\citep[][]{bergemann:08}. The obtained variations for lines of various 
multiplets often exceed 0.10 dex. Therefore, the LTE Mn abundances obtained 
from the lines of different multiplets must show systematic variations between 
them.
According to \cite{bergemann:08} due to the NLTE effects we should have obtained the 
systematic difference in LTE Mn abundance for two line groups of different 
multiplets.  We used pairs of the lines 4783--4823~\AA\ (multiplet 16, Elow =2.3 eV) and 
6013--6021~\AA\ (multiplet 27, Elow =3.07 eV). In Figs. \ref{el_fe1}, \ref{el_fe2} we show the dependences of 
the difference in abundances 
$\Delta Log A = Log Mn_{4783} - Log Mn_{6013}$
on the effective temperature and metallicity for the thin and thick disc stars.
There is no systematic trend observed. The average variations are 0.01$\pm$0.04 
and 0.02$\pm$0.04 for the thin and thick discs, respectively. Thus, the 
observations do not support the values of the LTE deviations obtained by 
\cite{bergemann:08} at the given temperatures and metallicities.

The development of an adequate model of Mn atoms to account for the effects of 
deviations from LTE is complicated by the absence of detailed computations of 
atomic data, such as photoionization cross-section or parameters of radiative 
and shock transitions. The use of approximations such as a H-like approximation 
for Mn atoms, yields NLTE corrections that are not robust. Taking all this into 
account, we believe instead that the LTE determinations for the Mn abundance 
are correct within the given uncertainty of 0.1 dex.

\section{Results and comparison with the literature.}
\label{sec: results_and_comparison}

\begin{figure} 
\begin{tabular}{c}
\includegraphics[width=8cm]{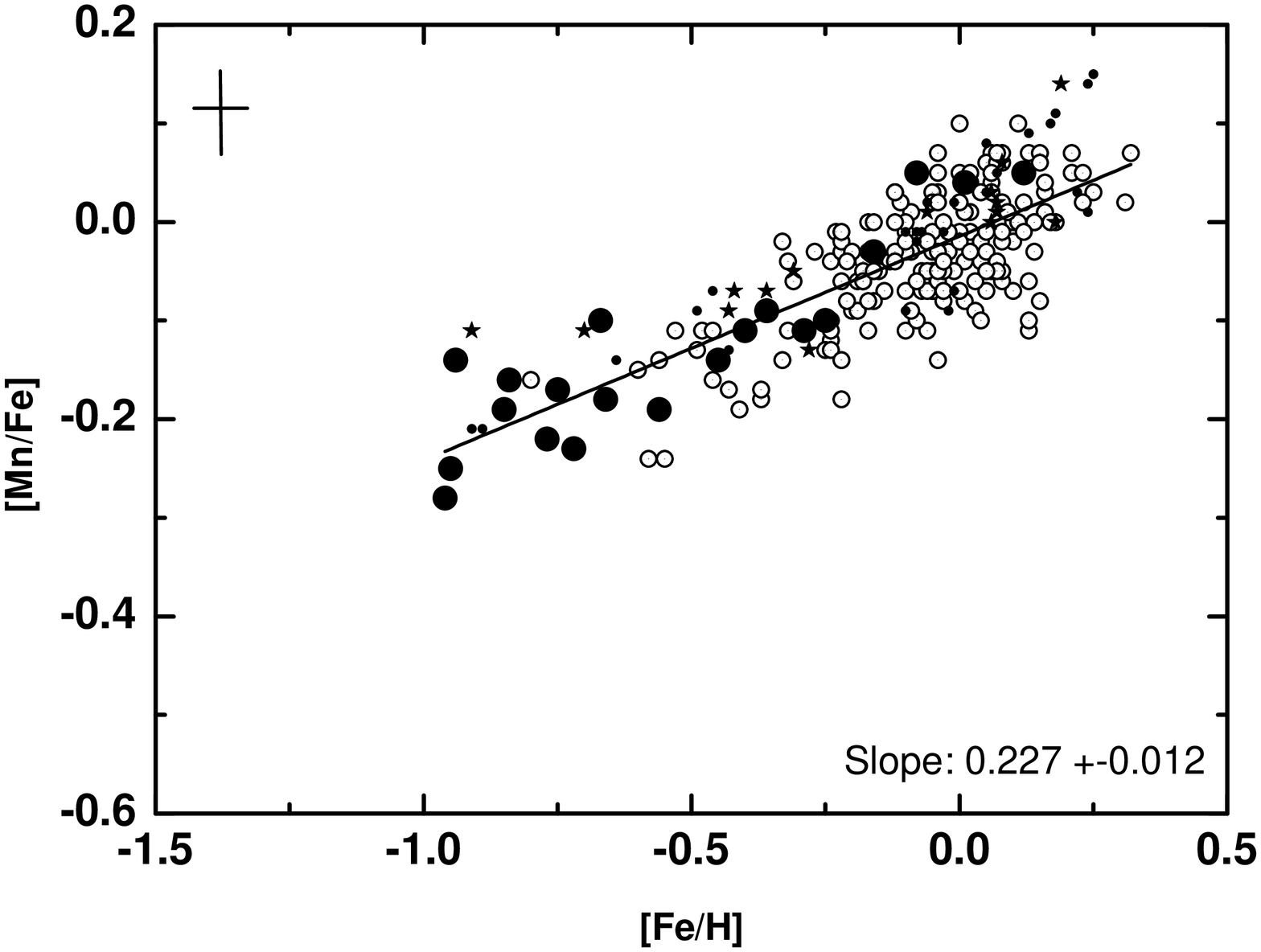}\\ 
\includegraphics[width=8cm]{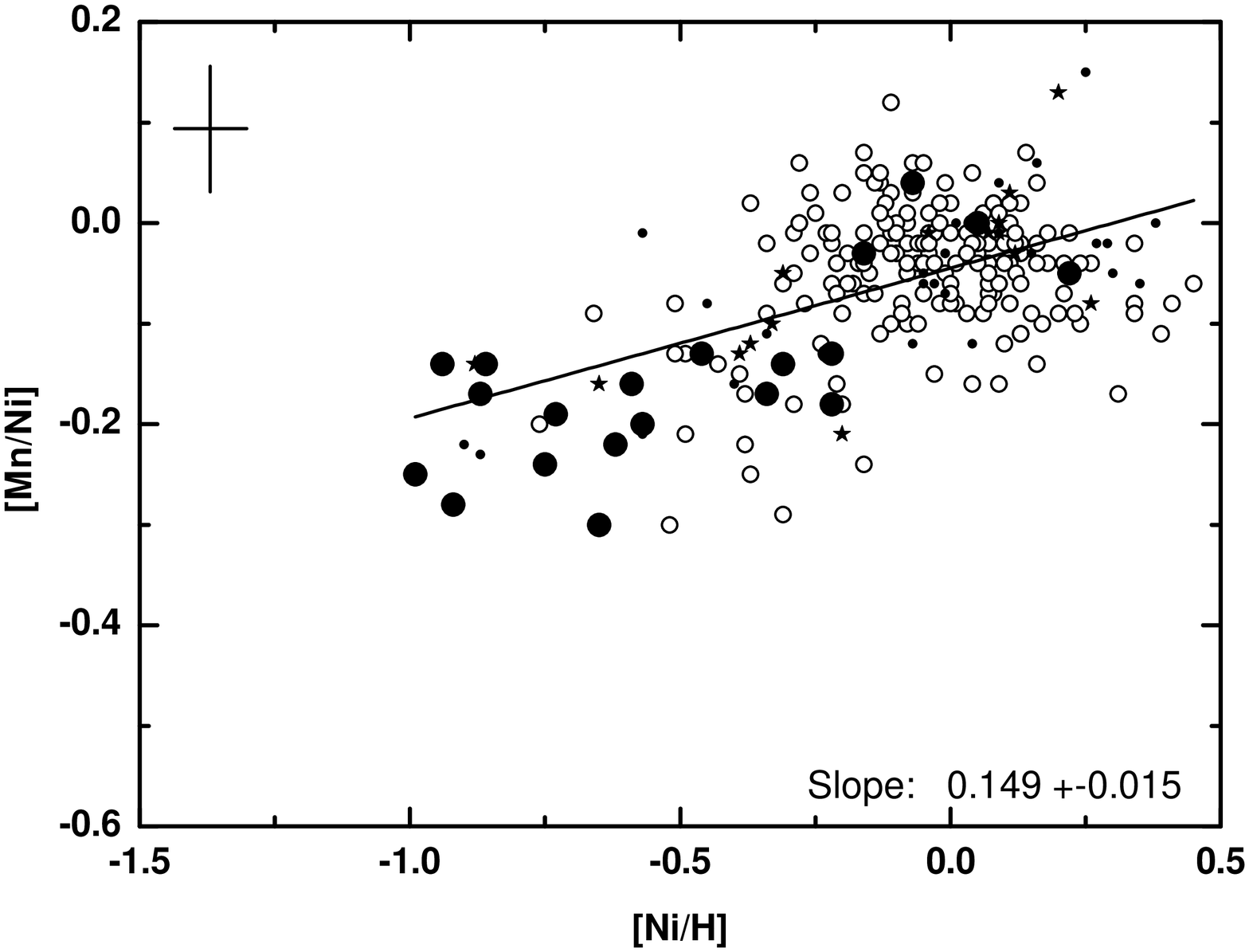}\\ 
\end{tabular}
\caption{
The trend of [Mn/Fe] ratio with respect to [Fe/H] is shown for our stellar 
sample: thin disc stars are marked as open symbols, thick disc stars as full 
symbols. Our determination for Hercules stream stars and unclassified stars are 
marked as black asterisks and points respectively. In the bottom figure we show 
the trend of [Mn/Ni] ratio with respect to [Ni/H]. 
}
\label{el_fe}
\end{figure}

\begin{figure*} 
\begin{tabular}{c}
\includegraphics[width=16cm]{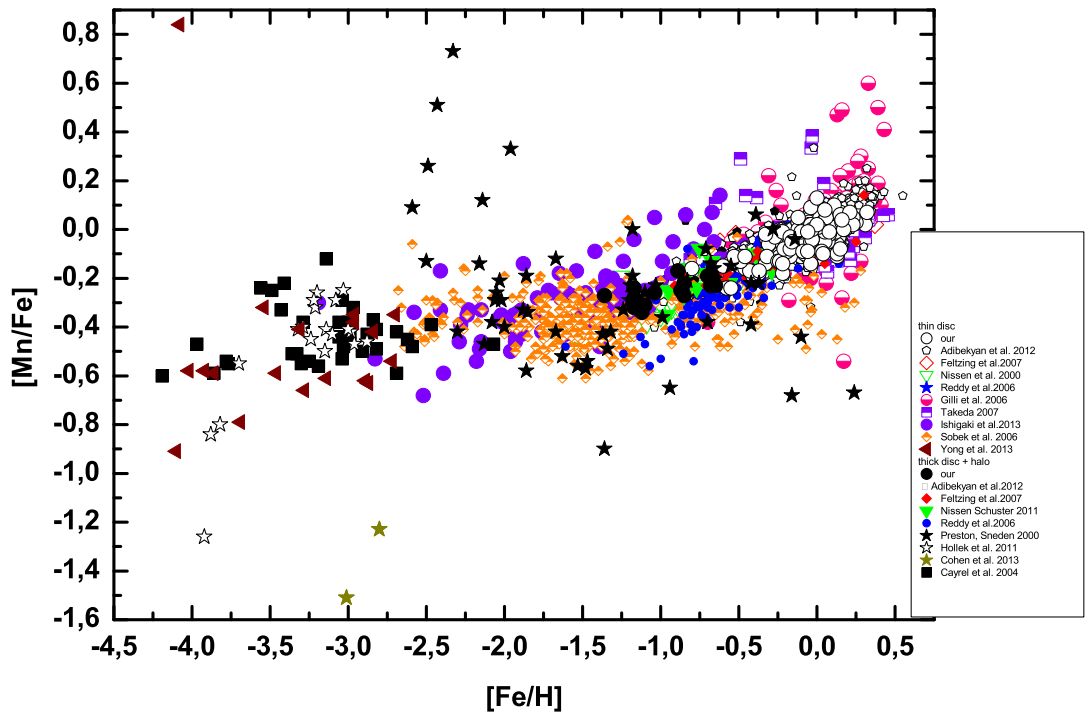}\\
\end{tabular}
\caption{
The trend of [Mn/Fe] ratio with respect to [Fe/H] is shown for our stellar 
sample, in comparison with the data of different authors. Markers are specified 
in the figure. 
}
\label{el_fe_comp}
\end{figure*}

The Mn abundances obtained for our stellar sample is shown in Fig. \ref{el_fe}, 
upper panel. In Figure \ref{el_fe_comp} our results are compared with other 
works for stars in the thin disc  \citep[][]
{adibekyan:12,nissen:00,  reddy:06,feltzing:07,gilli:06, takeda:07}, 
in the thick disc 
\citep[][]{adibekyan:12,nissen:11, reddy:06, feltzing:07, ishigaki:13}, in 
the Galactic halo \citep[][]{cayrel:04, ishigaki:13, preston:00, cohen:13, hollek:11,yong:13} and for
different populations \citep[][]{sobeck:06}. Part of the observational scatter 
is due to the use of different Mn lines and different methods for the analysis 
in the papers presented above. Different works adopted the LTE approximation, while 
\cite{bergemann:08} and \cite{battistini:15} used the NLTE approach. 
As we have shown above, our measurements for [Mn/Fe] are only marginally 
affected by the LTE assumption. Therefore, we are confident that we can also 
compare our results with these works. Most of the authors observe a similar 
increasing [Mn/Fe] trends with the increasing of the metallicity in the 
Galactic disc for $-1\lesssim$~[Fe/H]~$\lesssim 0.3$. \cite{takeda:07} did not 
observe any clear trend, but they used only one Mn line (5040~\AA). Stars from 
\cite{gilli:06} show a larger scatter at near-to-solar metallicity compared to 
other works: in particular, they obtain that stars hosting planetary systems 
show on average a larger Mn enrichment compared to stars without known planets. 

The Mn abundances obtained for the thick and thin disc stellar populations in 
with no difference between the thin disc and the thick disc 
within uncertainties. Stars in our sample belonging to the Hercules Stream show 
abundances consistent with the other Galactic disc populations. 
A detailed comparison between our results and the 
literature is given in Tables~\ref{compar}, and ~\ref{comparBB}. In particular, for stars in common 
with \cite{feltzing:07} we obtain similar results. The only exception is 
HD199960, where we obtain a [Mn/Fe] lower by 0.21 dex that is within the errors of the two works.
Note that the difference in the [Mn/Fe] for this star of our determination and the work of \cite{battistini:15} 
is 0.06 dex.  For other stars with Mn included in our sample and also measured by \cite{battistini:15}, 
the difference in [Mn/Fe] is within the determination error (0.10 dex) 
but for one star, HD 30495. In this case the difference is 0.15 dex, that is still consistent with our results 
within the errors given in BB 2015.

\section{Mn observations and discussion: implications from the nucleosynthesis of Mn in stars}
\label{sec: mn_nucleosynthesis}

In Fig. \ref{el_fe_comp} we have shown the evolution of Mn compared to Fe for 
stars at different metallicities, ranging from low-metallicity stars to 
super-solar metallicities. For stars with [Fe/H]~$\lesssim-1$, the abundance 
signature of Mn and Fe and therefore the observed [Mn/Fe] ratio is dominated by 
CCSNe \citep[e.g.,][]{kobayashi:11}, while for higher metallicities the 
influence of SNe type Ia dominates. The only stable isotope of Mn 
($^{55}$Mn) is made in explosive (complete and incomplete) Si-burning 
conditions as unstable $^{55}$Co, which decays afterwards via $^{55}$Fe to  
$^{55}$Mn. In complete Si-burning this production occurs under normal freeze-out 
conditions only for sufficiently high densities and/or low entropies 
\citep[see e.g.,][]{thielemann:86,thielemann:90}. 

Such conditions exist only in 
(near-)Chandrasekhar mass SNe type Ia 
\citep[][]{thielemann:86,iwamoto:99,brachwitz:00,seitenzahl:09,seitenzahl:13a,fink:14}. 
In the $\alpha$-rich regime of complete Si-burning (the only type of complete 
Si-burning experienced in CCSNe) $^{55}$Co abundance is moved over to $^{59}$Cu 
which decays via $^{59}$Ni to $^{59}$Co. However, $^{55}$Co is also produced in 
incomplete Si-burning, which takes place in CCSNe as well as SNe Ia 
\citep[][]{thielemann:86,woosley:95,thielemann:96,iwamoto:99,nakamura:99,brachwitz:00,woosley:02}. 
In all of these conditions, the production of $^{55}$Co depends also on the electron fraction ($Y_e$) of the matter experiencing explosive burning.  
$Y_e$ is the number of electrons per all nucleons (free and bound in nuclei), or 
the ratio of the number of all protons over all nucleons (i.e. neutrons plus protons). Thus, $Y_e$ = 0.5 indicates a stellar composition with equal numbers of neutrons and protons, $Y_e$ $<$ 0.5 means that it is neutron-rich and $Y_e$ $>$ 0.5 that is proton-rich. In stellar evolution, 
the electron fraction $Y_e$ first changes during H- and He-burning. Only 
marginal variations occur during the following C burning and Ne burning 
evolutionary stages. In advanced O-burning and then in Si-burning stages the  
$Y_e$ decreases significantly \citep[e.g.,][]{thielemann:85}. However, 
according to present theoretical stellar models these last regions will not be 
ejected by the SN explosion, or will not host the thermodynamic conditions 
needed to make Mn during the explosion. 

There exists, however, a $Y_e$ change in stellar models as a function of metallicity. 
In H-burning, the CNO-isotopes are burned essentially to $^{14}$N (an N=Z nucleus) 
which is moved in He-burning to 
$^{22}$Ne (an N=Z+2 nucleus due to the beta-decay of $^{18}$F to $^{18}$O during 
the $\alpha$-capture chain based on $^{14}$N). In this way the metallicity 
(given predominantly by CNO abundances) is turned into the abundance of 
$^{22}$Ne, which differs from the Z/A=0.5 of the other He-burning products and 
determines the electron fractions  $Y_e=\sum_i Z_i Y_i$. 
For lowest metallicities ([Fe/H]=-$\infty$) 
this relates to $Y_e=0.5$ after He-burning, for solar metallicities ([Fe/H]=0) 
to $Y_e=0.499$, for supersolar metallicities ([Fe/H]=0.25, 0.5) to  $Y_e=0.498, 
0.496$. $^{55}$Co, the radioactive progenitor of $^{55}$Mn, has a $Z/A=27/55$ of
0.491, which would be the  $Y_e$-value at which the highest $^{55}$Mn production
is expected, while low metallicity stars would not produce $^{55}$Mn if the 
$Y_e$ is only determined by initial metallicity.

\subsection{Massive stars}
\label{subsec: mn_ccsn}

In massive stars, as we mentioned above, Mn is made mostly by the SN explosion 
in incomplete explosive Si-burning conditions as $^{55}$Co 
\citep[e.g.,][]{woosley:95,thielemann:96,nakamura:99, woosley:02}. Mn production 
is increasing with the increase of the initial metallicity or, in other words, 
with the decrease of the electron fractions Y$_{\rm e}$ 
\citep[e.g.,][]{thielemann:96, nakamura:99, nomoto:13}.
Most of Fe is made as radiogenic $^{56}$Fe from radioactive $^{56}$Ni. 
This isotope is mainly made in complete explosive Si-burning conditions as 
primary product (i.e., independent of the initial metallicity of the star). 
However, a significant fraction is produced together with $^{55}$Co in less extreme 
(incomplete Si-burning) SN conditions, but with an increasing production with increasing Y$_{\rm e}$ (or decreasing of the initial metallicity), which 
is the opposite compared to $^{55}$Co. This scenario is becoming more 
complicated once the zoo of different CCSN types is considered. For instance, 
Hypernovae tend to produce lower [Mn/Fe] ratios compared to the less energetic 
SN Type-II, due to a larger production of $^{56}$Ni and more extended 
$\alpha$-rich freeze-out zones of complete Si-burning in comparison to incomplete 
Si-burning \citep[e.g.,][]{nakamura:99, nakamura:01, maeda:02}. Asymmetries 
before and after the CCSN, and the SN-shock propagation through the massive 
star progenitor \citep[e.g.,][]{thielemann:11,hix:14,wongwathanarat:15} will 
affect the final Mn/Fe ratio in the SN ejecta, possibly explaining the large 
[Mn/Fe] spread observed in the early Galaxy (Figure \ref{el_fe_comp}). 
Within this observational scatter, most of halo stars show an [Mn/Fe]$\sim-0.4$. 
Considering that CCSNe are producing about 30-50\% of the Fe observed in the 
Solar system, this means that 12-20\% of the solar Mn is made by CCSNe.  

\subsection{Type Ia supernovae}
\label{subsec: mn_snia}

For [Fe/H]~$\gtrsim-1$, the [Mn/Fe] ratio in disc stars is increasing up to the 
solar ratio, due to the contribution from SNe Ia \citep[SNIa, 
e.g.,][]{hillebrandt:13} with on average [Mn/Fe]$>$0
This is due to the fact that (near-)Chandrasekhar mass SNe type Ia experience also 
normal freeze-out conditions from nuclear statistical equilibrium in complete, 
explosive Si-burning, caused by high central densities and low entropies 
\citep[see e.g.,][]{thielemann:86,iwamoto:99,brachwitz:00,seitenzahl:09,seitenzahl:13a,fink:14}. 
In these inner high-density regions, the $Y_e$ is not due to the initial stellar 
metallicity (see introduction to this section), but caused by the capture of 
degenerate electrons with high Fermi energies on protons.  Even the reduction 
of theoretical electron-capture rates \citep[][]{brachwitz:00} did not change 
the amount of $^{55}$Co produced in these inner zones, because a region of 
similar mass content with the relevant $Y_e$ results also in that case. 
In particular, \cite{yamaguchi:15} recently have reported the first direct 
observation of high Mn/Fe ratios in the SNIa remnant 3C 397, that can be 
explained only by the low $Y_e$ due to electron captures. This makes 3C 397 an 
ideal candidate of a Chandrasekhar mass SNIa. In addition to these inner zones 
with normal freeze-out from complete Si-burning, also incomplete Si-burning is 
taking place in layers further out, where $Y_e$ is determined by the initial 
metallicity, i.e. where the production of Mn and Fe depends on the initial 
composition of the stellar progenitor. This results in a situation where the 
production of Mn and Fe in the inner zones is independent on the initial 
metallicity (just due to electron capture as a function of central density), 
while the production of Mn and Fe in the outer zones depends on the initial 
metallicity \citep[e.g.,][]{nomoto:84, thielemann:86, iwamoto:99, brachwitz:00,thielemann:03,seitenzahl:09, seitenzahl:13a, fink:14}. 
If we take e.g. the results of \citep[]{iwamoto:99} for the (near-)Chandrasekhar 
mass type Ia explosion model W7 based on progenitor stars of zero and solar 
metallicity ([Fe/H]=$-\infty$, 0), this leads to composition ratios in the 
ejecta of [Mn/Fe]=0.067 and 0.227, which would of course find their way into 
the ISM and new stars only after the appropriate delay times for their 
formation with that initial metallicity. By scaling the Mn production in outer 
layers varying linearly with metallicity (but keeping constant the Mn yields 
from the inner ejecta dominated by electron-capture) this would lead for 
[Fe/H]=0.25 and 0.5 to [Mn/Fe]=0.30 and 0.38, respectively. Other delayed 
detonation models find for solar metallicities values [Mn/Fe]=0.42 
\citep[see][]{seitenzahl:13}. All \mbox{(near-)Chandrasekhar} mass models lead to 
[Mn/Fe]$>0$, a contribution needed to explain the change from about $-0.4$ at 
low metallicities to [Mn/Fe]$>0.3$ at supersolar metallicities. 

Sub-Chandrasekhar mass type Ia models (see below) lack the inner 
electron-capture dominated Mn ejecta and contain only the outer metallicity 
dependent, incomplete Si-burning ejecta. Therefore, they will eject material 
with [Mn/Fe]$<0$. The reproduction of the observed [Mn/Fe] trend with respect 
to [Fe/H] in the Galactic disc is an important diagnostic for Galactic 
chemical evolution and the (type Ia) supernova models contributing to it. The 
increasing [Mn/Fe] trend has been considered as a signature: 1) of the gradual 
enrichment by SNIa ejecta of the ISM \citep[][]{kobayashi:09}; 2) of the 
increasing Mn/Fe ratio in the SNIa yields with the metallicity of the progenitor \citep[][]{cescutti:08}; 3) of the overlapping 
contribution of sub-Chandrasekhar SNe Ia made by WD mergers \citep[e.g.,][]{pakmor:10} or triggered by He-detonation on a 
single WD \citep[e.g.,][]{fink:10}, and SNe Ia reaching the Chandrasekhar mass by 
accretion on a WD \citep[e.g.,][]{thielemann:86, 
iwamoto:99, brachwitz:00,thielemann:03,seitenzahl:09, seitenzahl:13a, fink:14}. 
This last result has been discussed recently by \cite{seitenzahl:13}: 
sub-Chandrasekhar SNe Ia do not reach the conditions to make $^{55}$Co in nuclear 
statistical equilibrium opposite to more massive SNe Ia, yielding low Mn/Fe ratio ejecta. 
Therefore, the [Mn/Fe] trend observed in the galactic disc may be used as an 
indirect diagnostic of the relative contribution from different types of SNe Ia.
All the three arguments discussed above may play a role in defining the 
galactic trend of the [Mn/Fe] ratio, affecting in a similar way the evolution 
of the [Mn/Fe] with respect to [Fe/H]. The uncertainties associated with the 
nucleosynthesis of Mn and Fe in SNe Ia need to be also considered. While the 
nuclear uncertainties seem to be less relevant for this case 
\citep[][]{parikh:13}, other uncertainties associated with the SNIa explosion 
and to the stellar progenitor structure need to be considered. 

Mn and Fe have an opposite dependence on the metallicity of the SNIa progenitor, 
which makes the analysis more complicated. The contribution to the solar 
inventory by CCSNe and SNe Ia is quite similar for Ni and Fe, yielding a quite 
flat [Ni/Fe] for stars with metallicities lower than solar in the disc and in 
the galactic halo \citep[e.g.,][and references therein]{kobayashi:11,mishenina:13}.
As discussed in \cite{mishenina:13}, Ni is a primary product both in CCSNe and 
in SNe Ia, made by nuclear statistical equilibrium in both the two stellar 
sources 
\citep[see][for a recent analysis of Ni production compared to Fe in CCSN conditions]{jerkstrand:15}.
Therefore, for basic nucleosynthesis reasons the evolution of the 
[Mn/Ni] ratio should be an observational diagnostic for the production of Mn in 
stars much easier to study than the [Mn/Fe] ratio. 

We need to remind that the 
reproduction of the observed [Ni/Fe] trend in the galaxy has been proven to be 
challenging for GCE simulations, both in the halo and in the galactic disc 
\citep[e.g.,][]{goswami:00,kobayashi:11}. While assumptions made by GCE models 
may be an important source of uncertainty, the present issues to reproduce the 
[Ni/Fe] galactic trend is related to the present limitations in theoretical 
stellar models and, as a consequence, in the stellar yields used by GCE 
simulations. On the other hand, a confirmation of theoretical results comes 
for SNe Ia with close-to Chandrasekhar mass, with these objects yielding 
high Ni/Fe and low Mn/Ni ejecta \citep[][]{yamaguchi:15}. 
This means that the [Mn/Ni] ratio can be 
also used to distinguish different SNIa populations together with the [Mn/Fe] 
ratio, but without being affected by the metallicity dependence associated with
the Fe yields of SNe Ia. Consistent observations for Mn, Fe and Ni on the same 
stellar samples are important to study the production of Fe-group elements in 
SNe Ia. GCE studies aiming to deliver robust conclusions about the 
nucleosynthesis of Mn, should take into account both Fe and Ni as reference 
elements. 

In order to study the impact of this in our analysis, we also compare the Mn 
abundance with Ni (Fig. \ref{el_fe}). The average error for [Mn/Ni] is about 
0.15 dex \citep[][]{mishenina:13}. For our stellar sample, the average 
observed slope for [Mn/Fe] with respect to [Fe/H] is 0.227$\pm$0.012, while we 
obtain for [Mn/Ni] with respect to [Ni/H] 0.149$\pm$0.015. The two slopes are 
different. For the considerations made before, we may expect to observe a 
steeper slope for [Mn/Fe] compared to [Mn/Ni]. This confirm that the 
metallicity dependence of Fe yields in SNe Ia may play a role in the [Mn/Fe] and 
[Ni/Fe] trends. The impact of this with respect to the contribution from different 
SNIa populations to Mn, Fe and Ni still has to be investigated. At the moment 
we cannot derive any quantitative conclusion, since our thick disc sample does 
not include enough stars and because of observational errors. Furthermore, in 
our stellar sample we observe a larger dispersion of the [Mn/Ni] data compared 
to [Mn/Fe] in Fig. \ref{el_fe}, in particular for thin disc stars. 

\section{Conclusions and final remarks.}
\label{sec: conclusions}

In this work we presented and discussed the abundance measurements of Mn for 
247 disc stars. The analysis is based on LTE assumptions. 
The insufficient accuracy of atomic data makes difficult to construct an 
adequate model for NLTE calculations for Mn. We show that in our case the 
corrections by \cite{bergemann:08} are not confirmed by observations.
We have discussed the uncertainties affecting the determination of the Mn 
abundance. For [Mn/Fe] we obtain an error of about 0.10 dex.
The major source of its uncertainty is the choice of the temperature. 

For disc stars in our stellar sample we obtain an increasing [Mn/Fe] trend with 
[Fe/H] consistent with most of other works. Within observational uncertainties 
we cannot disentangle the abundance patterns for thin disc and thick disc stars, 
as obtained by \cite{feltzing:07, battistini:15}. 
On the other hand, our determinations of [Mn/Fe] are consistent with the data of these two 
works for common stars within the observational errors.             
We have compared the [Mn/Fe] and [Mn/Ni] trends with [Fe/H] and [Ni/H], 
respectively. The reason is that Mn and Fe production in SNe Ia both depend on 
the initial metallicity of the progenitor with opposite trends: Mn yields 
increase with the metallicity of the SNIa progenitor, while Fe yields decrease. 
On the other hand, Ni production is independent from the initial stellar 
metallicity. We show that the [Mn/Ni] and [Mn/Fe] patterns have an average 
slope of 0.149$\pm$0.015 and 0.227$\pm$0.012, respectively.
While the slopes are different within 2$\sigma$, the [Mn/Ni] observational 
dispersion for thin disc stars and our small sample of thick disc stars do not 
allow to derive quantitative conclusions.

We reviewed the production of Mn and Fe in SNe Ia and CCSNe.
In particular, there are three main scenarios that are qualitatively compatible 
with the observed [Mn/Fe] pattern in the galactic disc, including the relative 
contribution from both sub-Chandrasekhar mass SNe Ia and more massive SNe Ia. In 
order to define the relative frequency of the different SNIa populations 
explaining the [Mn/Fe] observations, the impact of the other two aspects need 
to be disentangled and weighted consistently by a detailed GCE study.

\section*{Acknowledgements}
TM, TG, MP, FKT and SK thank for the support from the Swiss National Science 
Foundation, project SCOPES No. IZ73Z0$_{}$152485.
MP acknowledges significant support to NuGrid from NSF grants PHY 09-22648 
(Joint Institute for Nuclear Astrophysics, JINA), NSF grant PHY-1430152 
(JINA Center for the Evolution of the Elements) and EU MIRG-CT-2006-046520.
MP acknowledges the support from the "Lendület-2014" Programme of the Hungarian 
Academy of Sciences (Hungary) and from SNF (Switzerland).
FKT acknowledges support from the European Research Council (FP7) under ERC 
Advanced Grant Agreement 321263 FISH.

\bibliography{2354m}

\appendix
\onecolumn

\begin{longtable}{lcccccccccc}
\caption{Atmospheric parameters and Mn abundance log A(Mn) each used line 
($\lambda$, A) and [Mn/Fe] ratio for our target stars}\\
\hline
Star & \Teff,K & \logg & \Vt & [Fe/H] & 4783, A  & 4823,A & 5432,A & 6013,A & 6021, A & [Mn/Fe] \\
\hline
\hline
\endfirsthead
\caption{Continued.} \\
\hline
Star & \Teff,K & \logg & \Vt & [Fe/H] & 4783, A  & 4823,A & 5432,A & 6013,A & 6021, A & [Mn/Fe] \\
\hline
\hline
\endhead
\hline
\endfoot
\hline
\endlastfoot
  Sun      &        &        &        &        &   5.3  &   5.25  &  5.27 &    5.24 &   5.24&        \\
Thick disk &        &        &        &        &        &         &       &         &       &        \\
HD245      &   5400 &   3.4  &  -0.84 &   0.7  &   4.32 &   4.35  &   4.1 &    4.18 &   4.18&  -0.16 \\
HD3765     &   5079 &   4.3  &   0.01 &   1.1  &        &         &   5.3 &    5.3  &   5.3 &   0.04 \\
HD6582     &   5240 &   4.3  &  -0.94 &   0.7  &   4.2  &   4.2   &  4.18 &    4.15 &   4.15&  -0.14 \\
HD13783    &   5350 &   4.1  &  -0.75 &   1.1  &   4.35 &   4.4   &   4.4 &         &   4.25&  -0.17 \\
HD18757    &   5741 &   4.3  &  -0.25 &    1   &   4.95 &   4.95  &  4.95 &    4.85 &   4.85&   -0.1 \\
HD22879    &   5972 &   4.5  &  -0.77 &   1.1  &   4.31 &   4.28  &       &    4.25 &   4.25&  -0.22 \\
HD65583    &   5373 &   4.6  &  -0.67 &   0.7  &   4.5  &   4.5   &  4.55 &    4.45 &   4.45&   -0.1 \\
HD76932    &   5840 &    4   &  -0.95 &    1   &   4.05 &   4.1   &       &    4.05 &   4.05&  -0.25 \\
HD106516   &   6165 &   4.4  &  -0.72 &   1.1  &   4.35 &   4.35  &       &         &   4.25&  -0.23 \\
HD110897   &   5925 &   4.2  &  -0.45 &   1.1  &   4.7  &   4.7   &  4.73 &    4.6  &   4.6 &  -0.14 \\
HD135204   &   5413 &    4   &  -0.16 &   1.1  &   5.15 &   5.05  &  5.08 &    5.05 &    5  &  -0.03 \\
HD152391   &   5495 &   4.3  &  -0.08 &   1.3  &   5.25 &   5.25  &  5.25 &    5.2  &   5.2 &   0.05 \\
HD157089   &   5785 &    4   &  -0.56 &    1   &   4.58 &   4.6   &  4.55 &    4.42 &   4.42&  -0.19 \\ 
HD157214   &   5820 &   4.5  &  -0.29 &    1   &   4.92 &   4.92  &  4.88 &    4.78 &   4.78&  -0.11 \\ 
HD159062   &   5414 &   4.3  &   -0.4 &    1   &   4.75 &   4.75  &   4.8 &    4.75 &   4.7 &  -0.11 \\ 
HD165401   &   5877 &   4.3  &  -0.36 &   1.1  &   4.85 &   4.85  &  4.85 &    4.75 &   4.75&  -0.09 \\ 
HD190360   &   5606 &   4.4  &   0.12 &   1.1  &        &         &  5.47 &    5.4  &   5.4 &   0.05 \\
HD201889   &   5600 &   4.1  &  -0.85 &   1.2  &   4.2  &   4.23  &   4.3 &    4.18 &   4.18&  -0.19 \\
HD201891   &   5850 &   4.4  &  -0.96 &    1   &   4.05 &   4.05  &       &     4   &   3.98&  -0.28 \\
HD204521   &   5809 &   4.6  &  -0.66 &   1.1  &   4.45 &   4.45  &   4.5 &    4.35 &   4.35&  -0.18 \\
Thin disk  &        &        &        &        &        &         &       &         &       &        \\
HD166      &   5514 &   4.6  &   0.16 &   0.6  &        &         &  5.42 &    5.42 &   5.42&   0.01 \\
HD1562     &   5828 &    4   &  -0.32 &   1.2  &   4.95 &   4.97  &       &    4.85 &   4.83&  -0.04 \\
HD1835     &   5790 &   4.5  &   0.13 &   1.1  &        &         &  5.45 &    5.45 &   5.45&   0.07 \\
HD3651     &   5277 &   4.5  &   0.15 &   0.6  &        &         &  5.45 &    5.45 &   5.5 &   0.07 \\
HD4256     &   5020 &   4.3  &   0.08 &   1.1  &        &         &  5.38 &    5.4  &   5.4 &   0.06 \\
HD4307     &   5889 &    4   &  -0.18 &   1.1  &   5.05 &   5.05  &  5.05 &    4.98 &   4.98&  -0.06 \\
HD4614     &   5965 &   4.4  &  -0.24 &   1.1  &        &         &    5  &    4.95 &   4.95&  -0.04 \\
HD5294     &   5779 &   4.1  &  -0.17 &   1.3  &   5.08 &   5.1   &  5.12 &    5.08 &   5.08&    0   \\
HD6660     &   4759 &   4.6  &   0.08 &   1.4  &        &         &   5.4 &    5.4  &   5.4 &   0.07 \\
HD7590     &   5962 &   4.4  &   -0.1 &   1.4  &   5.17 &   5.15  &  5.15 &    5.15 &   5.15&  -0.01 \\
HD7924     &   5165 &   4.4  &  -0.22 &   1.1  &        &         &    5  &     5   &    5  &  -0.03 \\
HD8648     &   5790 &   4.2  &   0.12 &   1.1  &        &         &   5.4 &    5.38 &   5.38&   0.02 \\
HD9407     &   5666 &   4.45 &   0.05 &   0.8  &   5.25 &   5.25  &  5.22 &    5.25 &   5.25&  -0.07 \\
HD9826     &   6074 &    4   &   0.1  &   1.3  &        &         &  5.27 &    5.3  &   5.27&  -0.07 \\
HD10086    &   5696 &   4.3  &   0.13 &   1.2  &        &         &  5.32 &    5.25 &   5.25&  -0.11 \\
HD10307    &   5881 &   4.3  &   0.02 &   1.1  &   5.3  &   5.28  &  5.28 &    5.25 &   5.22&  -0.01 \\
HD10476    &   5242 &   4.3  &  -0.05 &   1.1  &        &         &   5.2 &    5.12 &   5.12&  -0.05 \\
HD10780    &   5407 &   4.3  &   0.04 &   0.9  &   5.3  &   5.3   &  5.24 &    5.24 &   5.24&  -0.04 \\
HD11007    &   5980 &    4   &   -0.2 &   1.1  &    5   &    5    &    5  &    4.92 &   4.92&  -0.09 \\
HD11373    &   4783 &   4.65 &   0.08 &    1   &        &         &  5.35 &    5.25 &   5.25&  -0.05 \\
HD12846    &   5766 &   4.5  &  -0.24 &   1.2  &   4.92 &   4.92  &  4.94 &    4.87 &   4.84&  -0.12 \\
HD13507    &   5714 &   4.5  &  -0.02 &   1.1  &   5.22 &   5.22  &  5.26 &    5.15 &   5.18&  -0.03 \\
HD14374    &   5449 &   4.3  &  -0.09 &   1.1  &   5.24 &   5.24  &  5.18 &    5.13 &   5.13&   0.01 \\
HD16160    &   4829 &   4.6  &  -0.16 &   1.1  &        &         &  5.12 &    5.08 &   5.08&    0   \\
HD17674    &   5909 &    4   &  -0.14 &   1.1  &   5.12 &   5.12  &  5.05 &     5   &   4.98&  -0.07 \\
HD17925    &   5225 &   4.3  &  -0.04 &   1.1  &        &         &  5.25 &    5.3  &   5.3 &   0.07 \\
HD18632    &   5104 &   4.4  &   0.06 &   1.4  &        &         &  5.33 &    5.38 &   5.42&   0.07 \\
HD18803    &   5665 &   4.55 &   0.14 &   0.8  &        &         &   5.4 &    5.38 &   5.38&    0   \\
HD19019    &   6063 &    4   &  -0.17 &   1.1  &   5.08 &   5.08  &  5.05 &     5   &    5  &  -0.05 \\
HD19373    &   5963 &   4.2  &   0.06 &   1.1  &   5.4  &   5.4   &  5.36 &    5.32 &   5.32&   0.04 \\
HD20630    &   5709 &   4.5  &   0.08 &   1.1  &        &         &   5.3 &    5.25 &   5.25&  -0.06 \\
HD22484    &   6037 &   4.1  &  -0.03 &   1.1  &   5.22 &   5.2   &  5.22 &    5.15 &   5.15&  -0.04 \\
HD24053    &   5723 &   4.4  &   0.04 &   1.1  &   5.35 &   5.35  &       &    5.21 &   5.21&  -0.02 \\
HD24238    &   4996 &   4.3  &  -0.46 &    1   &        &         &  4.65 &    4.63 &   4.6 &  -0.16 \\
HD24496    &   5536 &   4.3  &  -0.13 &   1.5  &   5.08 &   5.08  &  5.18 &    5.08 &   5.04&  -0.04 \\
HD25665    &   4967 &   4.7  &   0.01 &   1.2  &        &         &  5.18 &    5.18 &   5.18&  -0.08 \\
HD25680    &   5843 &   4.5  &   0.05 &   1.1  &   5.3  &   5.3   &  5.28 &    5.25 &   5.25&  -0.03 \\
HD26923    &   5920 &   4.4  &  -0.03 &    1   &   5.2  &   5.2   &  5.12 &    5.13 &   5.13&  -0.07 \\
HD28447    &   5639 &    4   &  -0.09 &   1.1  &   5.15 &   5.15  &  5.22 &    5.1  &   5.1 &  -0.03 \\
HD29150    &   5733 &   4.3  &    0   &   1.1  &   5.31 &   5.33  &       &    5.18 &   5.18&  -0.01 \\
HD29310    &   5852 &   4.2  &   0.08 &   1.4  &        &         &  5.32 &    5.32 &   5.32&  -0.01 \\
HD29645    &   6009 &    4   &   0.14 &   1.3  &        &         &  5.38 &    5.35 &   5.35&  -0.03 \\
HD30495    &   5820 &   4.4  &  -0.05 &   1.3  &   5.2  &   5.2   &   5.2 &    5.15 &   5.15&  -0.03 \\
HD33632    &   6072 &   4.3  &  -0.24 &   1.1  &   4.95 &   4.95  &  4.95 &    4.88 &   4.88&   -0.1 \\
HD34411    &   5890 &   4.2  &   0.1  &   1.1  &        &         &  5.36 &    5.34 &   5.34&    0   \\
HD37008    &   5016 &   4.4  &  -0.41 &   0.8  &        &         &  4.68 &    4.64 &   4.62&  -0.19 \\
HD37394    &   5296 &   4.5  &   0.09 &   1.1  &        &         &  5.35 &    5.35 &   5.35&   0.01 \\
HD38858    &   5776 &   4.3  &  -0.23 &   1.1  &   5.05 &   5.05  &  5.07 &    4.97 &   4.97&  -0.01 \\
HD39587    &   5955 &   4.3  &  -0.03 &   1.5  &   5.2  &   5.2   &   5.2 &    5.1  &   5.05&  -0.08 \\
HD40616    &   5881 &    4   &  -0.22 &   1.1  &   5.08 &   5.08  &  5.08 &    4.92 &   4.92&  -0.02 \\
HD41330    &   5904 &   4.1  &  -0.18 &   1.2  &   5.08 &   5.08  &  5.08 &    4.98 &   4.98&  -0.04 \\
HD41593    &   5312 &   4.3  &  -0.04 &   1.1  &   5.3  &   5.3   &  5.23 &    5.2  &   5.2 &   0.03 \\
HD42618    &   5787 &   4.5  &  -0.07 &    1   &   5.15 &   5.15  &  5.18 &    5.08 &   5.08&  -0.06 \\
HD42807    &   5719 &   4.4  &  -0.03 &   1.1  &   5.22 &   5.25  &   5.2 &    5.15 &   5.15&  -0.04 \\
HD43587    &   5927 &   4.1  &  -0.11 &   1.3  &   5.2  &   5.2   &   5.2 &    5.15 &   5.12&   0.02 \\
HD43856    &   6143 &   4.1  &  -0.19 &   1.1  &   5.05 &   5.08  &  5.02 &    4.95 &   4.95&  -0.06 \\
HD43947    &   6001 &   4.3  &  -0.24 &   1.1  &   4.95 &   4.95  &  4.95 &    4.85 &   4.85&  -0.11 \\
HD47752    &   4613 &   4.6  &  -0.05 &   0.2  &        &         &  5.18 &    5.1  &   5.1 &  -0.07 \\
HD48682    &   5989 &   4.1  &   0.05 &   1.3  &   5.33 &   5.33  &  5.25 &    5.2  &   5.2 &  -0.05 \\
HD50281    &   4712 &   3.9  &   -0.2 &   1.6  &        &         &    5  &     5   &   5.05&  -0.03 \\
HD50692    &   5911 &   4.5  &   -0.1 &   0.9  &   5.08 &   5.08  &  5.08 &     5   &    5  &  -0.11 \\
HD51419    &   5746 &   4.1  &  -0.37 &   1.1  &        &   4.75  &       &    4.68 &   4.65&  -0.18 \\
HD51866    &   4934 &   4.4  &    0   &    1   &        &         &   5.3 &    5.3  &   5.3 &   0.05 \\
HD53927    &   4860 &   4.64 &  -0.22 &   1.2  &        &         &  4.85 &    4.85 &   4.85&  -0.18 \\
HD54371    &   5670 &   4.2  &   0.06 &   1.2  &   5.4  &   5.4   &  5.33 &    5.3  &   5.3 &   0.03 \\
HD58595    &   5707 &   4.3  &  -0.31 &   1.2  &   4.9  &   4.92  &       &    4.87 &   4.85&  -0.06 \\
HD59747    &   5126 &   4.4  &  -0.04 &   1.1  &        &         &  5.22 &    5.28 &   5.28&   0.05 \\
HD61606    &   4956 &   4.4  &  -0.12 &   1.3  &        &         &  5.18 &    5.15 &   5.15&   0.03 \\
HD62613    &   5541 &   4.4  &   -0.1 &   1.1  &   5.17 &   5.2   &  5.17 &    5.07 &   5.05&  -0.03 \\
HD63433    &   5693 &   4.35 &  -0.06 &   1.9  &   5.08 &   5.08  &  5.15 &    5.1  &   5.05&  -0.11 \\
HD63433    &   5693 &   4.35 &  -0.06 &   1.5  &   5.15 &   5.25  &   5.2 &    5.15 &   5.13&  -0.02 \\
HD64468    &   5014 &   4.2  &    0   &   1.2  &        &         &   5.4 &    5.32 &   5.32&   0.1  \\
HD64815    &   5864 &    4   &  -0.33 &   1.1  &   4.8  &   4.83  &  4.83 &    4.75 &   4.72&  -0.14 \\
HD65874    &   5936 &    4   &   0.05 &   1.3  &   5.4  &   5.4   &   5.4 &    5.34 &   5.32&   0.06 \\
HD72905    &   5884 &   4.4  &  -0.07 &   1.5  &   5.17 &   5.17  &       &    5.07 &   5.07&  -0.07 \\
HD73344    &   6060 &   4.1  &   0.08 &   1.1  &        &         &       &    5.27 &   5.27&  -0.05 \\
HD73667    &   4884 &   4.4  &  -0.58 &   0.9  &        &         &  4.45 &    4.45 &   4.4 &  -0.24 \\
HD91347    &   5931 &   4.4  &  -0.43 &   1.1  &   4.71 &   4.71  &       &    4.61 &   4.61&  -0.17 \\
HD101177   &   5932 &   4.1  &  -0.16 &   1.1  &   5.05 &   5.05  &  5.02 &    4.98 &   4.98&  -0.08 \\
HD102870   &   6055 &    4   &   0.13 &   1.4  &        &         &  5.35 &    5.3  &   5.3 &  -0.06 \\
HD105631   &   5416 &   4.4  &   0.16 &   1.2  &        &         &   5.4 &    5.4  &   5.45&   0.01 \\
HD107705   &   6040 &   4.2  &   0.06 &   1.4  &   5.3  &   5.3   &  5.25 &    5.25 &   5.25&  -0.05 \\
HD108954   &   6037 &   4.4  &  -0.12 &   1.1  &   5.15 &   5.15  &  5.15 &    5.05 &   5.05&  -0.03 \\
HD109358   &   5897 &   4.2  &  -0.18 &   1.1  &   5.08 &   5.08  &  5.05 &    4.98 &   4.98&  -0.05 \\
HD110463   &   4950 &   4.5  &  -0.05 &   1.2  &        &         &   5.2 &    5.12 &   5.12&  -0.05 \\
HD111395   &   5648 &   4.6  &   0.1  &   0.9  &        &         &  5.35 &    5.32 &   5.32&  -0.02 \\
HD112758   &   5203 &   4.2  &  -0.56 &   1.1  &        &         &  4.56 &    4.57 &   4.52&  -0.14 \\
HD114710   &   5954 &   4.3  &   0.07 &   1.1  &   5.35 &   5.35  &  5.25 &    5.25 &   5.25&  -0.04 \\
HD115383   &   6012 &   4.3  &   0.11 &   1.1  &        &         &  5.38 &    5.35 &   5.35&    0   \\
HD115675   &   4745 &   4.45 &   0.02 &    1   &        &         &  5.32 &    5.3  &   5.3 &   0.04 \\
HD116443   &   4976 &   3.9  &  -0.48 &   1.1  &        &         &   4.6 &    4.65 &   4.73&  -0.11 \\
HD116956   &   5386 &   4.55 &   0.08 &   1.2  &        &         &  5.35 &    5.35 &   5.35&   0.02 \\
HD117043   &   5610 &   4.5  &   0.21 &   0.4  &        &         &  5.52 &    5.5  &   5.5 &   0.05 \\
HD119802   &   4763 &    4   &  -0.05 &   1.1  &        &         &  5.25 &    5.2  &   5.25&   0.03 \\
HD122064   &   4937 &   4.5  &   0.07 &   1.1  &        &         &  5.45 &    5.35 &   5.35&   0.06 \\
HD124642   &   4722 &   4.65 &   0.02 &   1.3  &        &         &  5.35 &    5.3  &   5.3 &   0.05 \\
HD125184   &   5695 &   4.3  &   0.31 &   0.7  &        &         &  5.58 &    5.56 &   5.6 &   0.02 \\
HD126053   &   5728 &   4.2  &  -0.32 &   1.1  &   4.9  &   4.9   &       &    4.75 &   4.75&  -0.11 \\
HD127506   &   4542 &   4.6  &  -0.08 &   1.2  &        &         &   5.1 &    5.05 &   5.05&   -0.1 \\
HD128311   &   4960 &   4.4  &   0.03 &   1.3  &        &         &  5.22 &    5.22 &   5.22&  -0.06 \\
HD130307   &   4990 &   4.3  &  -0.25 &   1.4  &        &         &  4.85 &    4.88 &   4.88&  -0.13 \\
HD130948   &   5943 &   4.4  &  -0.05 &   1.3  &   5.18 &   5.18  &  5.18 &    5.08 &   5.08&  -0.07 \\
HD131977   &   4683 &   3.7  &  -0.24 &   1.8  &        &         &  4.85 &    4.93 &   4.96&  -0.13 \\
HD135599   &   5257 &   4.3  &  -0.12 &    1   &        &         &  5.15 &    5.13 &   5.2 &   0.03 \\
HD137107   &   6037 &   4.3  &    0   &   1.1  &   5.22 &   5.22  &  5.22 &    5.15 &   5.15&  -0.07 \\
HD139777   &   5771 &   4.4  &   0.01 &   0.7  &   5.28 &   5.28  &  5.22 &    5.18 &   5.18&  -0.04 \\
HD139813   &   5408 &   4.5  &    0   &   0.9  &   5.26 &   5.28  &  5.24 &    5.22 &   5.22&  -0.02 \\
HD140538   &   5675 &   4.5  &   0.02 &   0.9  &   5.32 &   5.32  &   5.3 &    5.25 &   5.25&   0.01 \\
HD141004   &   5884 &   4.1  &  -0.02 &   1.1  &   5.28 &   5.28  &  5.22 &    5.18 &   5.18&  -0.01 \\
HD141272   &   5311 &   4.4  &  -0.06 &   1.3  &   5.18 &   5.18  &  5.15 &    5.08 &   5.08&  -0.07 \\
HD142267   &   5856 &   4.5  &  -0.37 &   1.1  &   4.75 &   4.72  &  4.75 &    4.68 &   4.68&  -0.17 \\
HD144287   &   5414 &   4.5  &  -0.15 &   1.1  &   5.05 &   5.05  &   5.1 &    5.04 &   5.04&  -0.05 \\
HD145675   &   5406 &   4.5  &   0.32 &   1.3  &        &         &   5.7 &    5.63 &   5.6 &   0.07 \\
HD146233   &   5799 &   4.4  &   0.01 &   1.1  &   5.3  &   5.3   &   5.3 &    5.25 &   5.25&   0.01 \\
HD149661   &   5294 &   4.5  &  -0.04 &   1.1  &   5.22 &   5.22  &  5.18 &    5.18 &   5.15&  -0.03 \\
HD149806   &   5352 &   4.55 &   0.25 &   0.4  &        &         &  5.53 &    5.52 &   5.54&   0.03 \\
HD151541   &   5368 &   4.2  &  -0.22 &   1.3  &    5   &    5    &    5  &    4.95 &   4.95&  -0.06 \\
HD153525   &   4810 &   4.7  &  -0.04 &   1.0  &        &         &  5.18 &    5.13 &   5.13&  -0.06 \\
HD154345   &   5503 &   4.3  &  -0.21 &   1.3  &   5.05 &   5.08  &       &    4.97 &   4.95&  -0.04 \\
HD156668   &   4850 &   4.2  &  -0.07 &   1.2  &        &         &  5.08 &    5.15 &   5.15&  -0.05 \\
HD156985   &   4790 &   4.6  &  -0.18 &   1.0  &        &         &  5.02 &     5   &    5  &  -0.06 \\
HD158633   &   5290 &   4.2  &  -0.49 &   1.3  &        &         &   4.7 &    4.6  &   4.6 &  -0.13 \\
HD160346   &   4983 &   4.3  &   -0.1 &   1.1  &        &         &  5.15 &    5.15 &   5.15&    0   \\
HD161098   &   5617 &   4.3  &  -0.27 &   1.1  &   4.98 &   4.95  &       &    4.95 &   4.93&  -0.03 \\
HD164922   &   5392 &   4.3  &   0.04 &   1.1  &   5.32 &   5.32  &  5.35 &    5.32 &   5.32&   0.03 \\
HD165173   &   5505 &   4.3  &  -0.05 &   1.1  &   5.2  &   5.22  &  5.22 &    5.18 &   5.18&  -0.01 \\
HD165476   &   5845 &   4.1  &  -0.06 &   1.1  &   5.18 &   5.18  &   5.2 &    5.1  &   5.1 &  -0.05 \\
HD165670   &   6178 &    4   &   -0.1 &   1.5  &   5.12 &   5.12  &  5.12 &    5.05 &   5.05&  -0.07 \\
HD165908   &   5925 &   4.1  &   -0.6 &   1.1  &   4.55 &   4.55  &   4.4 &    4.48 &   4.45&  -0.15 \\
HD166620   &   5035 &    4   &  -0.22 &    1   &        &         &  4.95 &    5.05 &   5.05&  -0.01 \\
HD171314   &   4608 &   4.65 &   0.07 &   1.0  &        &         &   5.3 &    5.25 &   5.25&  -0.05 \\
HD174080   &   4764 &   4.55 &   0.04 &   1.0  &        &         &  5.32 &    5.32 &   5.32&   0.03 \\
HD175742   &   5030 &   4.5  &  -0.03 &   2.0  &        &         &  5.15 &    5.15 &   5.15&  -0.07 \\
HD176377   &   5901 &   4.4  &  -0.17 &   1.3  &    5   &    5    &  5.04 &    4.95 &   4.9 &  -0.11 \\
HD176841   &   5841 &   4.3  &   0.23 &   1.1  &        &         &  5.56 &    5.52 &   5.52&   0.05 \\
HD178428   &   5695 &   4.4  &   0.14 &   1.1  &        &         &  5.42 &    5.38 &   5.38&    0   \\
HD180161   &   5473 &   4.5  &   0.18 &   1.1  &        &         &  5.45 &    5.42 &   5.42&    0   \\
HD182488   &   5435 &   4.4  &   0.07 &   1.1  &        &         &  5.42 &    5.38 &   5.38&   0.07 \\
HD183341   &   5911 &   4.3  &  -0.01 &   1.3  &   5.2  &   5.2   &  5.22 &    5.18 &   5.18&  -0.05 \\
HD184385   &   5536 &   4.45 &   0.12 &   0.9  &        &         &  5.36 &    5.36 &   5.36&  -0.01 \\
HD185144   &   5271 &   4.2  &  -0.33 &   1.1  &        &         &   4.9 &    4.9  &   4.9 &  -0.02 \\
HD185414   &   5818 &   4.3  &  -0.04 &   1.1  &   5.13 &   5.1   &  5.08 &    5.06 &   5.03&  -0.14 \\
HD186408   &   5803 &   4.2  &   0.09 &   1.1  &        &         &  5.38 &    5.33 &   5.33&   0.01 \\
HD186427   &   5752 &   4.2  &   0.02 &   1.1  &   5.25 &   5.27  &  5.28 &    5.23 &   5.2 &  -0.03 \\
HD187897   &   5887 &   4.3  &   0.08 &   1.1  &        &         &  5.33 &    5.3  &   5.3 &  -0.02 \\
HD189087   &   5341 &   4.4  &  -0.12 &   1.1  &   5.16 &   5.16  &  5.18 &    5.1  &   5.1 &    0   \\
HD189733   &   5076 &   4.4  &  -0.03 &   1.5  &        &         &   5.2 &    5.15 &   5.15&  -0.05 \\
HD190007   &   4724 &   4.5  &   0.16 &   0.8  &        &         &  5.43 &    5.45 &   5.45&   0.03 \\
HD190406   &   5905 &   4.3  &   0.05 &   1.1  &   5.3  &   5.3   &  5.25 &    5.23 &   5.23&  -0.05 \\
HD190470   &   5130 &   4.3  &   0.11 &    1   &        &         &  5.47 &    5.45 &   5.45&   0.1  \\
HD190771   &   5766 &   4.3  &   0.13 &   1.5  &        &         &   5.3 &    5.28 &   5.25&   -0.1 \\
HD191533   &   6167 &   3.8  &   -0.1 &   1.5  &   5.2  &   5.2   &  5.15 &    5.08 &   5.08&  -0.02 \\
HD191785   &   5205 &   4.2  &  -0.12 &   1.2  &        &         &   5.1 &    5.08 &   5.08&  -0.04 \\
HD195005   &   6075 &   4.2  &  -0.06 &   1.3  &   5.21 &   5.2   &  5.21 &    5.15 &   5.15&  -0.02 \\
HD195104   &   6103 &   4.3  &  -0.19 &   1.1  &   5.02 &   5.02  &  4.95 &    4.95 &   4.95&  -0.09 \\
HD197076   &   5821 &   4.3  &  -0.17 &   1.2  &   5.08 &   5.08  &  5.08 &    5.02 &   5.02&  -0.03 \\
HD199960   &   5878 &   4.2  &   0.23 &   1.1  &        &         &   5.5 &    5.5  &   5.5 &   0.02 \\
HD200560   &   5039 &   4.4  &   0.06 &   1.1  &        &         &  5.36 &    5.36 &   5.36&   0.05 \\
HD202108   &   5712 &   4.2  &  -0.21 &   1.1  &   4.98 &   4.98  &    5  &    4.95 &   4.95&  -0.08 \\
HD202575   &   4667 &   4.6  &  -0.03 &   0.5  &        &         &   5.2 &    5.17 &   5.17&  -0.04 \\
HD203235   &   6071 &   4.1  &   0.05 &   1.3  &   5.33 &   5.33  &   5.3 &    5.28 &   5.28&  -0.01 \\
HD205702   &   6020 &   4.2  &   0.01 &   1.1  &   5.32 &   5.3   &   5.3 &    5.25 &   5.22&   0.01 \\
HD208038   &   4982 &   4.4  &  -0.08 &    1   &        &         &  5.07 &    5.12 &   5.15&  -0.06 \\
HD208313   &   5055 &   4.3  &  -0.05 &    1   &        &         &   5.2 &    5.22 &   5.25&   0.02 \\
HD208906   &   5965 &   4.2  &   -0.8 &   1.7  &   4.3  &   4.35  &       &         &   4.25&  -0.16 \\
HD210667   &   5461 &   4.5  &   0.15 &   0.9  &        &         &  5.45 &    5.45 &   5.48&   0.06 \\
HD210752   &   6014 &   4.6  &  -0.53 &   1.1  &   4.65 &   4.68  &  4.65 &    4.58 &   4.55&  -0.11 \\
HD211472   &   5319 &   4.4  &  -0.04 &   1.1  &   5.18 &   5.22  &  5.17 &    5.12 &   5.17&  -0.05 \\
HD214683   &   4747 &   4.6  &  -0.46 &   1.2  &        &         &       &    4.7  &   4.65&  -0.11 \\
HD216259   &   4833 &   4.6  &  -0.55 &   0.5  &        &         &       &    4.45 &   4.45&  -0.24 \\
HD216520   &   5119 &   4.4  &  -0.17 &   1.4  &        &         &  4.95 &    5.02 &   5.02&  -0.08 \\
HD217014   &   5763 &   4.3  &   0.17 &   1.1  &        &         &  5.42 &    5.42 &   5.42&    0   \\
HD217813   &   5845 &   4.3  &   0.03 &   1.5  &   5.2  &   5.2   &  5.25 &    5.17 &   5.17&  -0.09 \\
HD218868   &   5547 &   4.45 &   0.21 &   0.4  &        &         &  5.52 &    5.54 &   5.54&   0.07 \\
HD219538   &   5078 &   4.5  &  -0.04 &   1.1  &        &         &  5.18 &    5.25 &   5.25&   0.02 \\
HD219623   &   5949 &   4.2  &   0.04 &   1.2  &   5.2  &   5.2   &  5.18 &    5.2  &   5.2 &   -0.1 \\
HD220140   &   5144 &   4.6  &  -0.03 &   2.4  &        &         &  5.18 &    5.18 &   5.18&  -0.04 \\
HD220182   &   5364 &   4.5  &  -0.03 &   1.2  &   5.25 &   5.25  &  5.25 &    5.18 &   5.2 &    0   \\
HD220221   &   4868 &   4.5  &   0.16 &   0.5  &        &         &  5.53 &    5.38 &   5.45&   0.04 \\
HD221851   &   5184 &   4.4  &  -0.09 &    1   &        &         &  5.07 &    5.05 &   5.1 &  -0.09 \\
HD222143   &   5823 &   4.45 &   0.15 &   1.1  &        &         &  5.37 &    5.28 &   5.32&  -0.08 \\
HD224465   &   5745 &   4.5  &   0.08 &   0.8  &        &         &  5.35 &    5.3  &   5.32&  -0.01 \\
HD263175   &   4734 &   4.5  &  -0.16 &   0.5  &        &         &  5.08 &    5.02 &   5.02&  -0.05 \\
HDBD+12063 &   4859 &   4.4  &  -0.22 &   0.6  &        &         &  4.85 &    4.88 &   4.94&  -0.14 \\
HDBD+124499&   4678 &   4.7  &    0   &   0.5  &        &         &  5.32 &    5.25 &   5.25&   0.02 \\
Hercules   &        &        &        &        &        &         &       &         &       &        \\
HD13403    &  5724  &   4    & -0.31  &  1.1   &  4.88  &  4.88   & 4.93  &   4.9   &  4.9  &   -0.05 \\
HD19308    &  5844  &  4.3   &  0.08  &  1.1   &        &         & 5.44  &   5.36  &  5.36 &    0.06 \\
HD23050    &  5929  &  4.4   & -0.36  &  1.1   &  4.87  &  4.85   & 4.87  &   4.8   &  4.77 &   -0.07 \\
HD30562    &  5859  &   4    &  0.18  &  1.1   &        &         & 5.44  &   5.41  &  5.44 &     0   \\
HD64606    &  5250  &  4.2   & -0.91  &  0.8   &  4.26  &  4.26   & 4.28  &   4.2   &  4.2  &   -0.11 \\
HD68017    &  5651  &  4.2   & -0.42  &  1.1   &  4.78  &  4.78   & 4.83  &   4.73  &  4.73 &   -0.07 \\
HD81809    &  5782  &   4    & -0.28  &  1.3   &  4.85  &  4.85   & 4.92  &   4.82  &  4.8  &   -0.13 \\
HD107213   &  6156  &  4.1   &  0.07  &  1.6   &  5.35  &  5.35   & 5.35  &   5.35  &  5.3  &    0.01 \\
HD139323   &  5204  &  4.6   &  0.19  &  0.7   &        &         & 5.65  &   5.55  &  5.55 &    0.14 \\
HD144579   &  5294  &  4.1   &  -0.7  &  1.3   &  4.45  &  4.45   & 4.53  &   4.4   &  4.4  &   -0.11 \\
HD159222   &  5834  &  4.3   &  0.06  &  1.2   &  5.38  &  5.38   & 5.37  &   5.33  &  5.28 &    0.03 \\
HD159909   &  5749  &  4.1   &  0.06  &  1.1   &  5.36  &  5.36   & 5.32  &   5.28  &  5.28 &     0   \\
HD215704   &  5418  &  4.2   &  0.07  &  1.1   &        &         & 5.38  &   5.32  &  5.32 &    0.02 \\
HD218209   &  5705  &  4.5   & -0.43  &   1    &  4.75  &  4.75   & 4.79  &   4.72  &  4.69 &   -0.09 \\
HD221354   &  5242  &  4.1   & -0.06  &  1.2   &        &         &  5.2  &   5.2   &  5.2  &    0.01 \\
 Nonclass  &        &        &        &        &        &         &       &         &       &         \\
HD4628     &  4905  &  4.6   & -0.36  &  0.5   &        &         &  4.8  &   4.8   &  4.8  &   -0.09 \\
HD4635     &  5103  &  4.4   &  0.07  &  0.8   &        &         &  5.4  &   5.35  &  5.35 &    0.05 \\
HD10145    &  5673  &  4.4   & -0.01  &  1.1   &  5.18  &  5.18   & 5.22  &   5.15  &  5.15 &   -0.07 \\
HD12051    &  5458  &  4.55  &  0.24  &  0.5   &        &         &  5.5  &   5.5   &  5.5  &    0.01 \\
HD13974    &  5590  &  3.8   & -0.49  &  1.1   &  4.7   &  4.7    &  4.7  &   4.65  &  4.65 &   -0.09 \\
HD17660    &  4713  &  4.75  &  0.17  &  1.3   &        &         & 5.55  &   5.5   &  5.5  &    0.1  \\
HD20165    &  5145  &  4.4   & -0.08  &  1.1   &        &         & 5.18  &   5.15  &  5.15 &   -0.01 \\
HD24206    &  5633  &  4.5   & -0.08  &  1.1   &  5.2   &  5.18   & 5.18  &   5.12  &  5.1  &   -0.02 \\
HD32147    &  4945  &  4.4   &  0.13  &  1.1   &        &         &  5.5  &   5.45  &  5.45 &    0.09 \\
HD45067    &  6058  &   4    & -0.02  &  1.2   &  5.2   &  5.2    & 5.17  &   5.1   &  5.1  &   -0.09 \\
HD84035    &  4808  &  4.8   &  0.25  &  0.5   &        &         &  5.7  &   5.62  &  5.62 &    0.15 \\
HD86728    &  5725  &  4.3   &  0.22  &  0.9   &        &         &  5.5  &   5.5   &  5.5  &    0.03 \\
HD90875    &  4788  &  4.5   &  0.24  &  0.5   &        &         & 5.72  &   5.62  &  5.62 &    0.14 \\
HD117176   &  5611  &   4    & -0.03  &   1    &  5.22  &  5.25   & 5.25  &   5.18  &  5.18 &   -0.01 \\
HD117635   &  5230  &  4.3   & -0.46  &  0.7   &        &         & 4.75  &   4.7   &  4.7  &   -0.07 \\
HD154931   &  5910  &   4    &  -0.1  &  1.1   &  5.17  &  5.18   & 5.18  &   5.1   &  5.1  &   -0.01 \\
HD159482   &  5620  &  4.1   & -0.89  &   1    &  4.2   &  4.2    &       &   4.12  &  4.12 &   -0.21 \\
HD168009   &  5826  &  4.1   & -0.01  &  1.1   &  5.3   &  5.3    &  5.3  &   5.25  &  5.22 &    0.02 \\
HD173701   &  5423  &  4.4   &  0.18  &  1.1   &        &         & 5.58  &   5.52  &  5.52 &    0.11 \\
HD182736   &  5430  &  3.7   & -0.06  &   1    &  5.26  &  5.26   & 5.28  &   5.16  &  5.16 &    0.02 \\
HD184499   &  5750  &   4    & -0.64  &  1.5   &  4.5   &  4.5    &  4.5  &   4.45  &  4.45 &   -0.14 \\
HD184768   &  5713  &  4.2   & -0.07  &  1.1   &  5.2   &  5.2    &  5.2  &   5.15  &  5.15 &   -0.01 \\
HD186104   &  5753  &  4.2   &  0.05  &  1.1   &  5.38  &  5.38   &       &   5.3   &  5.3  &    0.03 \\
HD215065   &  5726  &   4    & -0.43  &  1.1   &  4.75  &  4.75   &       &   4.65  &  4.65 &   -0.13 \\
HD219134   &  4900  &  4.2   &  0.05  &  0.8   &        &         & 5.39  &   5.35  &  5.39 &    0.08 \\
HD219396   &  5733  &   4    &  -0.1  &  1.2   &  5.12  &  5.14   &       &   5.03  &  4.98 &   -0.09 \\
HD224930   &  5300  &  4.1   & -0.91  &  0.7   &  4.2   &  4.2    &  4.1  &   4.1   &  4.1  &   -0.21 \\
\hline                                                                                               
\end{longtable}

\label{lastpage}

\bsp

\end{document}